\documentclass[useAMS,usenatbib]{mn2e}
\usepackage{times}
\usepackage{graphicx}

\newcommand{\etal}{{et~al.}\,}      
\newcommand{\eg}{{e.g.,}\,}         
\newcommand{\ie}{{i.e.,}\,}         
\newcommand{\MSUN}{${\rm M}_\odot$}
\newcommand{\Msun}{\,{\rm M}_\odot}
\newcommand{\iraf}{\textsc{iraf}}

\DeclareRobustCommand{\ion}[2]{%
\relax\ifmmode
\ifx\testbx\f
{\mathrm{#1\,\textsc{#2}}}\else
{\mathrm{#1\,\mathsc{#2}}}\fi
\else\textup{#1\,{\mdseries\textsc{#2}}}%
\fi}

\def\apj    {{ApJ }}
\def\apjs   {{ApJS }}
\def\apss   {{ApSS }}
\def\aj     {{AJ }}
\def\aas    {{A\&AS }}
\def\aaa    {{A\&A }}
\def\pasp   {{PASP }}
\def\mnras  {{MNRAS }}

\title[NIR Photometric Properties of Six Local Volume Dwarf Galaxies]{Photometric Properties of Six Local Volume Dwarf Galaxies from Deep Near-Infrared Observations}
\author[B. de Swardt, R. C. Kraan-Korteweg and H. Jerjen]{B. de Swardt$^{1,2}$\thanks{E-mail:
bonita@saao.ac.za}, R. C. Kraan-Korteweg$^{2}$ and H. Jerjen$^{3}$ \\
$^{1}$South African Astronomical Observatory, PO Box 9, Observatory 7935, South Africa. \\
$^{2}$Department of Astronomy, University of Cape Town, Rondebosch 7701, South Africa. \\
$^{3}$Research School of Astronomy and Astrophysics, Australian National University, Mount Stromlo Observatory, Cotter Road, \\ Weston ACT 2611, Australia.}

\begin{document}


\pagerange{\pageref{firstpage}--\pageref{lastpage}} \pubyear{2009}

\maketitle

\label{firstpage}

\begin{abstract}
We have obtained deep near-infrared $J$- ($\rm 1.25\mu m$), $H$- ($\rm 1.65\mu m$) and $K_s$-band ($\rm 2.15\mu m$) imaging for a sample of six dwarf galaxies ($M_B \ga -17~\rm mag$) in the Local Volume (LV, $D\la10\rm~Mpc$). The sample consists mainly of early-type dwarf galaxies found in various environments in the LV. Two galaxies (LEDA~166099 and UGCA~200) in the sample are detected in the near-infrared for the first time. The deep near-infrared images allow for a detailed study of the photometric and structural properties of each galaxy. The surface brightness profiles of the galaxies are detected down to the $\sim24~\rm mag~arcsec^{-2}$ isophote in the $J$- and $H$-bands, and $23~\rm mag~arcsec^{-2}$ in the $K_s$-band. The total magnitudes of the galaxies are derived in the three wavelength bands. For the brightest galaxies ($M_B\la-15.5~\rm mag$) in the sample, we find that the Two Micron All Sky Survey (2MASS) underestimates the total magnitudes of these systems by up to $\la0.5~\rm mag$. The radial surface brightness profiles of the galaxies are fitted with an exponential (for those galaxies having a stellar disk) or S\'ersic law to derive the structure of the underlying stellar component. In particular, the effective surface brightness ($\mu_e$) and effective radius ($r_e$) are determined from the analytic fits to the surface brightness profile. The $J$-$K_s$ colours for the galaxies have been measured to explore the luminosity-metallicity relation for early-type dwarfs. In addition, the $B$-$K_s$ colours of the galaxies are used to assess their evolutionary state relative to other galaxy morphologies. The total stellar masses of the dwarf galaxies are derived from the $H$-band photometric measurements. These will later be compared to the dynamical mass estimates for the galaxies to determine their dark matter content.      
\end{abstract}


\section{Introduction}

Dwarf galaxies are, apart from stellar clusters, the most fundamental stellar systems. They are believed to be the building blocks of larger galaxies in hierarchical galaxy formation theories \citep[\eg][]{Bul2001} and can provide insight into galaxy formation and evolution on the smallest scales \citep{Tay2003}. The Local Volume (LV, $D\la10\rm~Mpc$) provides a diverse environment in which these systems can be studied. The observations of dwarf galaxies in the LV, however, remain challenging due to their characteristic low surface brightness ($\mu_{V, \rm eff}\ga22 \rm~mag~arcsec^{-2}$). The availability of more sensitive detectors has led to the discovery of numerous dwarf galaxies in the LV. These stellar systems can now be studied in unprecedented detail to measure their physical properties (\eg stellar, gas and dark matter content). Overall, an understanding of the physical mechanisms driving the observational properties of dwarf galaxies can significantly impact our current views of galaxy formation and evolution, as well as the nature of dark matter. 

\begin{table*}
 \centering
 \begin{minipage}{160mm}
  \caption{Basic properties of target dwarf galaxies. Columns: (1) Galaxy designation; (2) Morphological classification taken from \citet{Jer2000b} and \citet{Par2002}, with the exception of NGC~5206 which is given by the NASA/IPAC Extragalactic Database (NED); (3) and (4) Equatorial coordinates for J2000 epoch; (5) and (6) Distance to the galaxy in Megaparsecs with an indication of the method used \citep{Jer2000a,Jer2000b,Kara2004}: SBF -- via fluctuation of surface brightness; MEM -- from membership in the known groups; (7) Heliocentric radial velocity of galaxy: $^a$Using the optical emission lines in the spectrum of NGC~59 \citep{Bea2006}, $^b$The heliocentric velocity of UGCA~200 was computed by averaging the radial velocities of six globular clusters measured by \citet{Puz2008}. Note that the globular cluster having an elevated radial velocity of $v_{\odot}=1210\pm27\rm km~s^{-1}$ was excluded as it might be associated with the disk component of NGC~3115 rather than UGCA~200; $^c$\citet{daC98}; $^d$\citet{Cot97}; $^e$\citet{Jer2000a}; (8) Total apparent $B$-band magnitude; (9) Absolute $B$-band magnitude (corrected for Galactic extinction); (10) Reddening estimate from \citet{Sch1998}; (11) $B\mbox{-}R$ colour. The photometric parameters were taken from \citet{Par2002} and \citet{Jer2000b}, with the exception of NGC~5206 which is taken from \citet{Lau89}. \label{dwarf_properties1} }
  \begin{tabular}{@{}lcccccccccc@{}}
  \hline
         &	  & \multicolumn{2}{c}{J2000} &  $D$ 	&	   & $v_{\odot}$         & {$B$} & $M_B$ & $E(B\mbox{-}V)$ & $B\mbox{-}R$ \\
{Galaxy} & {Type} & RA & DEC                  & {(Mpc)} & {Method} & {($\rm km~s^{-1}$)} & {(mag)} & {(mag)} & {(mag)}  & {(mag)} \\
(1) & (2) & (3) & (4) & (5) & (6) & (7) & (8) & (9) & (10) & (11) \\
  \hline
NGC~59 	      & dS0     & 00 15 25.1 & $-$21 26 40 & 4.4 & SBF & $362\pm10~^a$ & 12.97 & -15.74 & 0.020 & 1.04 \\
LEDA~166099   & dE,N    & 09 12 29.3 & $-$24 14 28 & 9.8 & MEM & -- 	       & 15.50 & -14.46 & 0.197 & 1.16 \\
UGCA~200      & dE,N    & 10 05 35.1 & $-$07 45 00 & 9.7 & MEM & $721\pm30~^b$ & 16.16 & -13.78 & 0.048 & 1.38 \\
NGC~3115 DW01 & dE,N    & 10 05 41.6 & $-$07 58 53 & 9.7 & MEM & $698\pm42~^c$ & 13.38 & -16.55 & 0.052 & 1.38 \\
NGC~5206      & SB?(r?) & 13 33 44.0 & $-$48 09 04 & 3.6 & MEM & $571\pm10~^d$ & 11.64 & -16.66 & 0.120 & 1.22  \\
ESO~384-016   & dS0/Im  & 13 57 01.4 & $-$35 19 59 & 4.2 & SBF & $561\pm32~^e$ & 15.11 & -13.06 & 0.074 & 0.71  \\
\hline
\end{tabular}
\end{minipage}
\end{table*}

Extensive optical galaxy surveys of the local neighbourhood ($D\la10\rm Mpc$) of the Milky Way \citep{Kra1979,Sch1992,Kara2004} have been carried out for well over two decades. Increasingly fainter and lower surface brightness (SB) galaxies have been detected within this time frame. These include the discovery of numerous dwarf galaxies in the LV \citep{Cot97,Kar1998,Jer1998,Karaseva1999,Karaseva2000,Kara2000,Jer2000a,Kara2001}. Follow-up observations of the newly discovered dwarfs have primarily focused on obtaining distance estimates and stellar population studies for these galaxies \citep[see][and references therein]{Kara1999,Kar2007,Sei2001}. Deep $B$- and $R$-band imaging of the new dwarf members in the Centaurus~A and Sculptor groups was performed by \citet{Jer1998,Jer2000a,Jer2000b}. They derive a distance estimate for each galaxy using the surface brightness fluctuation method \citep{Ton1988}. The photometric and structural parameters of the galaxies were also measured. A detailed $BR$ photometric analysis of a larger sample of LV dwarf galaxies (from the \citet{Kar1998} sample) was later carried out by \citet{Par2002}. Their goal was to establish a surface photometry database for a large homogeneous sample of nearby dwarf galaxies. The photometric database provides a systematic means of studying structural differences in the dwarf galaxies. Multi-wavelength observations ranging from the optical to the infrared wavelengths are, however, needed to gain a complete understanding of the morphology and evolutionary state of these galaxies. 

The main contribution to the stellar mass of galaxies arises from the underlying old stellar component. The morphology of the individual galaxies is established by this underlying component which is thought to be the ``backbone" of the galaxy. The distribution of the old stellar component is effectively probed at near-infrared (NIR) wavelengths. The NIR wavelengths are minimally influenced by dust attenuation revealing the internal galaxy structure. The largest NIR database containing the photometric and structural parameters of nearby dwarf galaxies has been constructed using observations from the Two Micron All Sky Survey \citep[2MASS,][]{Jar2000,Jar2003}. The inadequacy of the 2MASS photometry for deriving these parameters is, however, continuously being highlighted as deeper observations of the dwarf galaxies are obtained. These effects are more evident for the low SB galaxies where the short exposures of 2MASS result in either the galaxy being undetected or its total flux being underestimated by up to 70\% \citep{And2002,Kir2008}. Deep NIR observations of the LV dwarf galaxies are thus needed to avoid the selection effects of 2MASS photometry at low luminosities and low star densities. 

Deep NIR observations are hardly available for nearby dwarf galaxies because of the large integration times required for imaging these faint stellar systems. The deep NIR imaging of nearby dwarfs have subsequently focused on those galaxies showing star formation activity such as the dwarf irregular galaxies \citep{Vad05} and blue compact dwarfs \citep[\eg][]{Cai03}. Recently, \citet{Kir2008} have obtained deep $H$-band ($1.65\mu\rm m$) observations of a large sample of 57 LV galaxies consisting mostly of irregular galaxies. Their deep observations allow for the galaxies to be detected $4\rm~ mag~arcsec^{-2}$ or 40 times fainter than 2MASS. Given the high spatial resolution of the images, they were able to derive photometric and structural parameters for the galaxies. 

The observations by \citet{Kir2008} and \citet{Vad05} are seen as the first contributions to an extensive NIR photometric database of LV dwarf galaxies. To add to this effort, we have obtained simultaneous deep $J$- ($\rm 1.25\mu m$), $H$- ($\rm 1.65\mu m$) and $K_s$-band ($\rm 2.15\mu m$) observations of a sample of six LV dwarf galaxies. The galaxies are members of a larger sample of $\sim40$ LV dwarf galaxies for which deep NIR observations will eventually be obtained. The photometric analysis includes the derivation of the radial SB profiles and the colour profiles of the galaxies (section~\ref{sec5:SBprofiles}). The total magnitudes of the galaxies are derived in section~\ref{Total_mag}. Here it can be seen that the deep $JHK_s$ observations allow for a more accurate measure of the galaxy magnitudes compared to 2MASS. The structural parameters of the underlying stellar component are obtained in section~\ref{sec6:SBfits} by fitting either an exponential or S\'ersic law to the radial SB profile of the galaxy. The NIR photometric parameters of the three faintest dwarfs ($M_B\ga-15.5\rm~mag$) in the sample are measured for the first time. A detailed discussion of the photometric results for the six dwarfs is given in section~\ref{sec7:NIRdiscussion}. In addition, we also derived the NIR luminosities and total stellar masses of the galaxies (section~\ref{NIRmass}). The NIR photometric results presented in this paper will complement the $BV$ surface photometry database already established for these galaxies. 

\section[]{Sample Selection}

Deep near-infrared (NIR) $J$-, $H$- and $K_s$-band imaging was obtained for six Local Volume (LV, $D\la10~\rmn{Mpc}$) dwarf galaxies. The following criteria led to the selection of the target galaxies for NIR observations:

\begin{enumerate}
\renewcommand{\theenumi}{(\arabic{enumi})}
\item The dwarf galaxies are required to have an angular size smaller than $\sim$4\arcmin\ in the $B$-band so that the galaxy falls completely in the 7\farcm8$\times$7\farcm8 field-of-view (FoV) of the IRSF telescope. This criterion will allow a reliable measure of the sky background within the FoV of the galaxy images.   

\item Given the wide range in right ascension of all the target galaxies in our large sample, the galaxies within the February--June visibility window of the South African Astronomical Observatory (SAAO) site were selected. This was the period corresponding to the allocated semester for observing the dwarf galaxies. 

\item Extremely low surface brightness (SB) galaxies ($\mu_{B,\rmn{eff}}\ga24.5 \rmn{~mag~arcsec^{-2}}$) were as yet excluded. 

\end{enumerate}

The basic properties of the six target galaxies are listed in Table~\ref{dwarf_properties1}. The sample consists mainly of early-type dwarf galaxies spanning a range of distances ($3.5$--$10$ Mpc) within the LV. The galaxies ESO~384-016 and NGC~5206 were identified as part of the Centaurus~A (Cen A) group \citep{Cot97,Jer2000a}, while NGC~59 was found to be a member of the Sculptor (Scl) group \citep{Jer1998}. The remaining galaxies in the sample are more distant dwellers and lie on the outskirts of the LV \citep[and references therein]{Kara2004}. The galaxies NGC~59, NGC~3115~DW01 and NGC~5206 are the brightest in the sample with total apparent $B$-band magnitude of $B\la13.4~\rm mag$. The fainter dwarf galaxies in the sample ($B > 15~\rm mag$) are less spatially extended on the sky with LEDA~166099 having an angular size of $D_{\rm ext}<1$ arcminute. 

\section[]{Observations} \label{chap3:DataAcqu}

Deep NIR imaging of the six dwarf galaxies was obtained with the 1.4m Infrared Survey Facility\footnote{A description of the telescope can be found at http://www.z.phys.nagoya-u.ac.jp/$\sim$telescop/index\_e.html~.} (IRSF) telescope in Sutherland, South Africa \citep{Gla2000}. The \textit{Sirius} detector of the IRSF telescope consists of three 1024$\times$1024 HgCdTe arrays. The array system allows for simultaneous 3-channel $J$- ($\rm 1.25\mu m$), $H$- ($\rm 1.65\mu m$) and $K_s$-band ($\rm 2.15\mu m$) imaging. The individual CCDs give a total FoV of 7\farcm8$\times$7\farcm8 together with a 0\farcs45 pixel scale. 

The NIR data were obtained during three different observing runs over the period of 2006--2007. A log of all the galaxy observations is presented in Table~\ref{NIRobs1}. The observations performed in 2006 February and June were carried out using shared telescope time. Single galaxy observations of NGC~3115~DW01 and NGC~59 were obtained during these respective observing runs. A further week in 2007 March was dedicated to the imaging of dwarf and low SB galaxies. Simultaneous $J$-, $H$- and $K_s$-band imaging of the other four galaxies was obtained during this observing run. The total integration times for the galaxy observations vary from 60--132 minutes. All observations were performed during grey time under photometric conditions. 

The aim of the NIR observations is to conduct a detailed photometric analysis of the dwarf galaxies out to at least the $\mu_{K_s}\sim23~\rm~mag~arcsec^{-2}$ isophote. R. Metcalfe \& M. McCall (private communication) derived an exposure time for detecting nearby dwarf irregular (dIrr) galaxies down to this isophote using the IRSF telescope. The exposure time was calculated based on NIR observations of these galaxies with the 3.6m Canada-France-Hawaii telescope (CFHT) and the 2.1m OAN-SPM telescope \citep{Vad05}. They found that a limiting magnitude of $\mu_{K_s}\sim23~\rm~mag~arcsec^{-2}$ can be reached in $\sim$70 minutes. The goal was therefore to observe each galaxy in principle for at least 70 minutes. An exception was made for the luminous galaxy NGC~59 which was only observed for 60 minutes on the shared night of June 11 2006. The low SB dwarf galaxies, LEDA~166099 and UGCA~200, were both observed for 96~minutes to ensure their detection (see Table~\ref{NIRobs1}).    

The quality of deep NIR images is greatly affected by temporal and spatial variations in the sky background. A good estimate of the sky-level is particularly important when observing low SB dwarf galaxies such as LEDA~166099 and UGCA~200. The NIR observing technique of \citet{Vad04} was employed to ensure an optimal extraction of the sky-level in the images. They propose an observing sequence of the form:  \begin{equation} \rm{sky-galaxy-sky-~....~-sky-galaxy-sky}~, \label{eq:ObsSeq} \end{equation} when observing faint extended sources in the NIR. Temporal variations in the sky-level were accounted for by allowing equal exposures for the galaxy and sky frames. A dithering step of 10\arcsec\ was applied to each new galaxy and sky exposure in the observing sequence. This was necessary for the removal of bad pixels and contaminants from the images. 

\begin{table}
 \centering
 \caption{Observing Log
   \label{NIRobs1}}
  \begin{tabular}{@{}lrcccc@{}}
  \hline
       &        &         & Exp & Total Exp \\
Galaxy & {Date (UT)} & Filters & (min) & (min) \\ 
\hline
 NGC~3115~DW01 & 2006 Feb 14 & $JHK_s$ &  48 & 132\\
               &       Feb 16    & $JHK_s$ &  24 & \\
               &       Feb 17    & $JHK_s$ &  60 & \\
 NGC~59        & 2006  Jun 11 & $JHK_s$ &  60 & 60 \\
 LEDA~166099   & 2007  Mar 08 & $JHK_s$ &  96 & 96 \\
 ESO~384-016   & 2007  Mar 08 & $JHK_s$ &  24 & 84 \\
	       &       Mar 09     & $JHK_s$ &  60  & \\
 UGCA~200      & 2007  Mar 10 & $JHK_s$ &  24 & 96 \\
	       &       Mar 11     & $JHK_s$ &  24 & \\
	       &       Mar 13     & $JHK_s$ &  48 & \\
 NGC~5206      & 2007  Mar 10 & $JHK_s$ &  48 & 60 \\
	       &       Mar 13     & $JHK_s$ &  12 & \\
\hline
\end{tabular}
\end{table}
 
The sequential sky and galaxy frames in (\ref{eq:ObsSeq}) were exposed for 60 seconds each. Individual exposures were sub-divided into 3$\times$20s non-dithered frames to avoid saturation of the pixel arrays. The background was sampled by choosing a sky region in close proximity to the galaxy that shows the least amount of stellar contamination. The sky-level was sampled either 10\arcmin\ North or South of the galaxy center.   

Twilight sky images were obtained in the evening and morning. A sequence of equal-duration exposures was taken as the twilight brightened or faded in each filter. These sky flats were used to remove the spatial variations in the images. A series of dark exposures were taken every morning in the $J$-, $H$-, and $K_s$-bands for removal of the detector signature.  

\section[]{Data Reduction and Calibration} \label{chap3:Reduction}

The data reduction was carried out using standard tasks in \iraf\footnote{\iraf\ is distributed by the National Optical Astronomy Observatory (NOAO), which is operated by the Association of Universities for Research in Astronomy, Inc., under cooperative agreement with the National Science Foundation.}. The first step in the reduction procedure involves the removal of the bias-level from the target images (\ie sky flats, galaxy and sky exposures). Dark frames having the same exposure time as the target image were used to create a master dark frame which was then subtracted from the individual target exposures. A master flat was created in each filter from a sequence of 30--40 equal-duration twilight sky exposures. The master flat was used to correct for the pixel-to-pixel sensitivity in the sky and galaxy exposures. 

The non-dithered 20sec galaxy and sky exposures were combined giving an observational sequence of the form: \begin{equation} S_1 - T_1 - S_2 - ... - S_i - T_i - S_{i+1} - ... - S_n - T_n - S_{n+1}, \label{sky3} \end{equation} where a dithered region of the sky $S_i$ is sampled before and after each galaxy exposure $T_i$. This technique of straddling the sky frames allows for the interpolation of the background at the time of the galaxy observation. The sky subtraction procedure of \citet{Vad04} was followed in subtracting the background from the galaxy images. Basically, this procedure creates a smooth background by first subtracting adjacent sky frames using two different combinations \ie $S_i - S_{i+1}$ and $S_{i+1} - S_i$. The background level at the time of the galaxy observation is obtained by averaging the resulting sky frames. The background level and corresponding uncertainty were measured at various locations in the galaxy image. These measurements were carried out at a radius of $r\ga4\arcmin$ from the galaxy center to ensure that the results are not influenced by light from the galaxy itself. We find that the reduction method described above removes the sky to an accuracy of $\sim$0.02\% in the $K_s$-band relative to the original signal.

\begin{table}
 \centering
 \caption{Properties of the reduced NIR images
   \label{NIRobs2}}
  \begin{tabular}{@{}llcc@{}}
\hline
        &          & {Exp} & {Ave seeing} \\ 
 {Galaxy} & {Filter} & {(min)} & {(arcsec)} \\
\hline
 NGC 3115 DW01 & $J$   &  57 & 1.3 \\
               & $H$   &  57 & 1.2 \\
               & $K_s$ &  57 & 1.2 \\
 NGC 59 & $J$   &  56 & 1.6 \\
        & $H$   &  57 & 1.5 \\
        & $K_s$ &  58 & 1.4 \\
 LEDA 166099 & $J$   & 89 & 1.4 \\
             & $H$   & 93 & 1.3 \\
             & $K_s$ & 94 & 1.3 \\
 ESO 384-016 & $J$   & 76 & 1.5 \\
             & $H$   & 83 & 1.5 \\
             & $K_s$ & 83 & 1.4 \\
 UGCA 200 & $J$   & 74  & 1.5 \\
          & $H$   & 73  & 1.4 \\
          & $K_s$ & 81  & 1.4 \\
 NGC 5206 & $J$   & 44  & 1.3 \\
          & $H$   & 44  & 1.3 \\
          & $K_s$ & 47  & 1.2 \\
\hline
\end{tabular}
\end{table}

The total on-source exposure time of the reduced galaxy images is listed in Table \ref{NIRobs2}. Those galaxy exposures largely affected by atmospheric and telescopic defects were not used in creating the final images. This results in the different exposure times in the $J$-, $H$- and $K_s$-bands for a single galaxy observation. Defects in the images include extreme blurring due to changes in the observing conditions as well as problems with the CCD readout. The seeing in the final galaxy images varies from 1\farcs2 in the $K_s$-band to 1\farcs6 in the $J$-band. 

The reduced $K_s$-band galaxy images are shown in Fig.~\ref{NIR_im1}. A distinct nucleus is seen in the galaxies NGC~3115 DW01, NGC~59, LEDA~166099 and NGC~5206. The brighter galaxies NGC~3115 DW01, NGC~59 and NCG~5206 show extended light profiles at least 1\arcmin\ beyond the nuclear component. The white features seen for example in the $K_s$-band image of LEDA~166099 are negative residuals left behind by the sky subtraction. These features are observed in the crowded stellar fields where it is more challenging to obtain a smooth sky background. 

\begin{figure} 
    \begin{center} 
    \includegraphics[width=6cm]{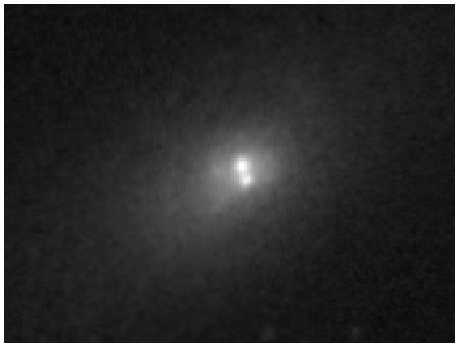}
	\caption[Double nuclear component of the Scl group dwarf galaxy NGC~59 in the NIR.]{Double nuclear component of the Scl group dwarf galaxy NGC~59 as seen in the $H$-band. The two peaks are separated by $\sim$2\farcs3. The image size shown is 68\arcsec$\times$58\arcsec. North is up and East is left.}
	\label{n59zoom}
\end{center}
\end{figure}

The central region of the galaxy NGC~59 is shown in Fig.~\ref{n59zoom}. Two intensity peaks aligned in the northeast-southwest direction can be distinguished in the center of this galaxy. The two peaks are separated by $\sim$2\farcs3 with the northern component being more luminous in all three wavelength bands. Recent star formation activity has been detected in the center of NGC~59 by \citet{Ski03}.  In this paper, we assume that the northern component is the ``true" nucleus of the galaxy while the second component is a star-forming region. The photometry of the two nuclear components together with a detailed kinematic study of NGC~59, will be presented in a companion paper (de Swardt et al., in preparation).

\subsection[]{Photometric Calibration}

The photometric calibrations of the IRSF data involve a direct comparison of the instrumental magnitudes of point sources in the field to their corresponding apparent magnitudes given by 2MASS. A colour correction was applied to the instrumental magnitudes to account for differences in the IRSF and 2MASS filter systems. The point sources used to calibrate the galaxy images were selected from the 2MASS Point Source Catalogue \citep{Skr06} and were chosen to satisfy the following criteria:
\begin{enumerate}
\item{Point sources should be brighter than the 2MASS completeness limit: $J\la15.8\rm~mag$, $H\la15.1\rm~mag$ and $K_s\la14.3\rm~mag$.}

\item{`AAA' quality photometry is available for all stellar sources in the 2MASS catalogue. Point sources are rated as having `AAA' quality photometry if they have a magnitude uncertainty of less than 10\% in all three wavelength bands.}

\item{Saturated stellar sources having magnitudes brighter than $10\rm~mag$ in the IRSF images were not used in the photometric calibrations.} 
\end{enumerate}

The instrumental magnitudes of the point sources were transformed to apparent magnitudes using the equations:
\begin{equation} j = J + j_1 + j_2(J - K_s)~, \label{jcalib} \end{equation}   
\begin{equation} h = H + h_1 + h_2(J - H)~, \label{hcalib} \end{equation}   
\begin{equation} k_s = K_s + k_1 + k_2(J - K_s)~, \label{kcalib} \end{equation}   
where $J$, $H$ and $K_s$ are the apparent magnitudes of the stars. The instrumental magnitudes are given by $j$, $h$ and $k_s$ in the $J$-, $H$- and $K_s$-bands, respectively. The nightly zero-points in the different wavelength bands are $j_1$, $h_1$ and $k_1$. These were determined for the individual galaxy images of each night. The offset between the instrumental and apparent magnitude (given by the 2MASS Point Source catalogue) was calculated for each of the point sources. The uncertainty in the magnitude offset is given by the square of the internal errors associated with the 2MASS and instrumental magnitudes, respectively. The magnitude offsets of the point sources show little scatter ($< 0.05\rm~mag$) so that the nightly zero-point was taken as the mean magnitude offset between the IRSF and 2MASS point sources.

\begin{figure*}
    \includegraphics[width=6.5cm]{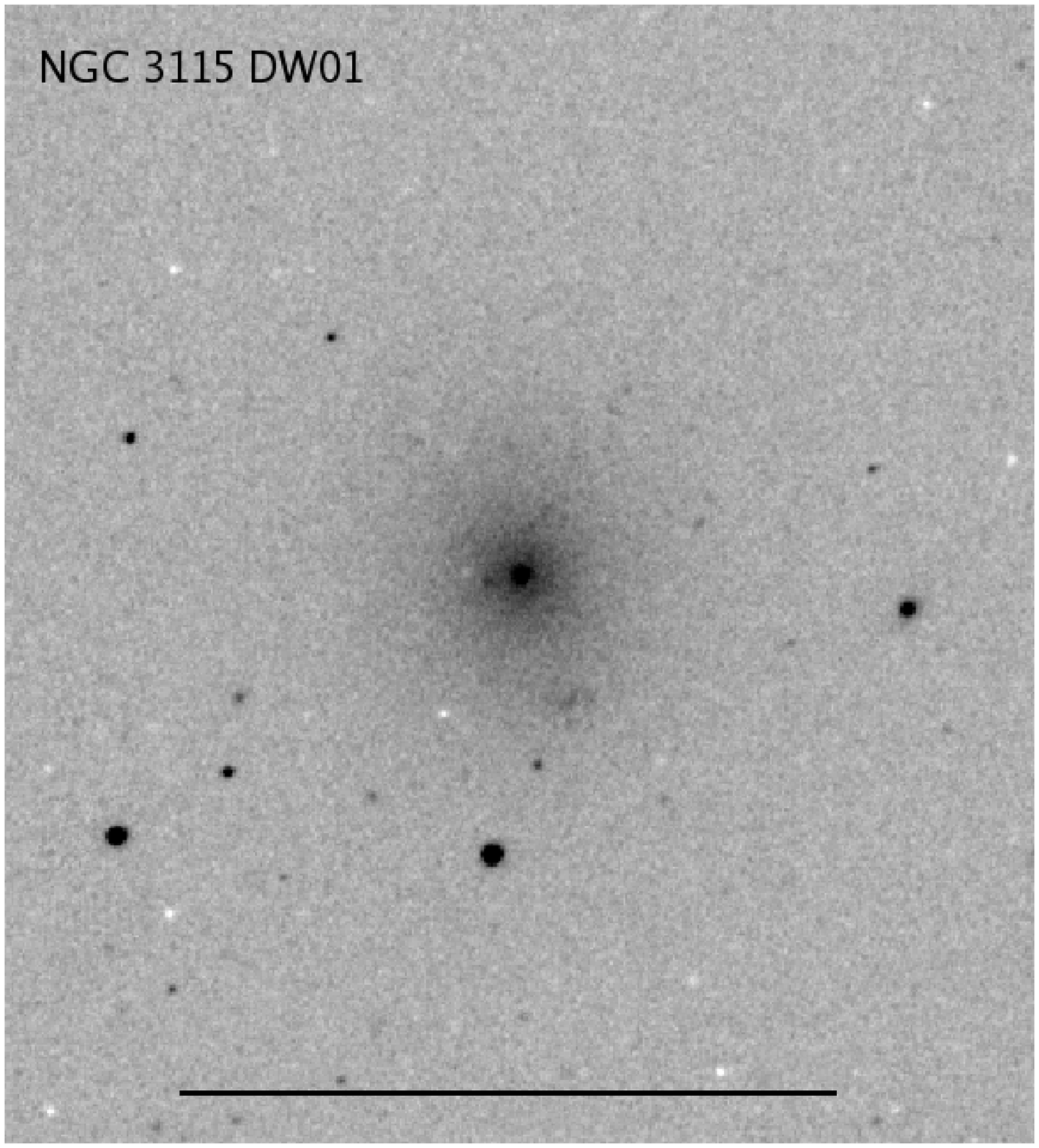}
    \includegraphics[width=7.5cm]{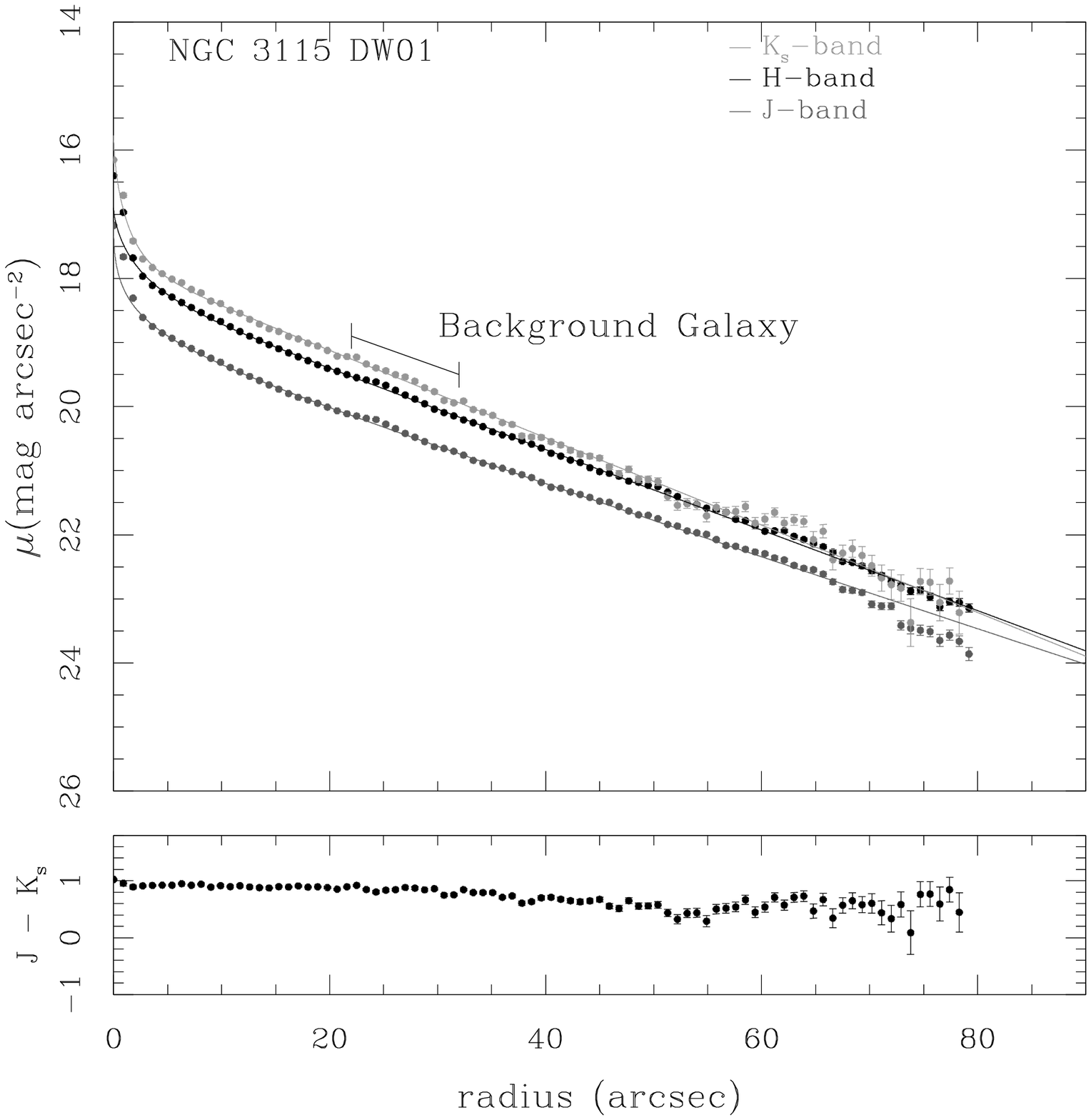}
    \includegraphics[width=6.5cm]{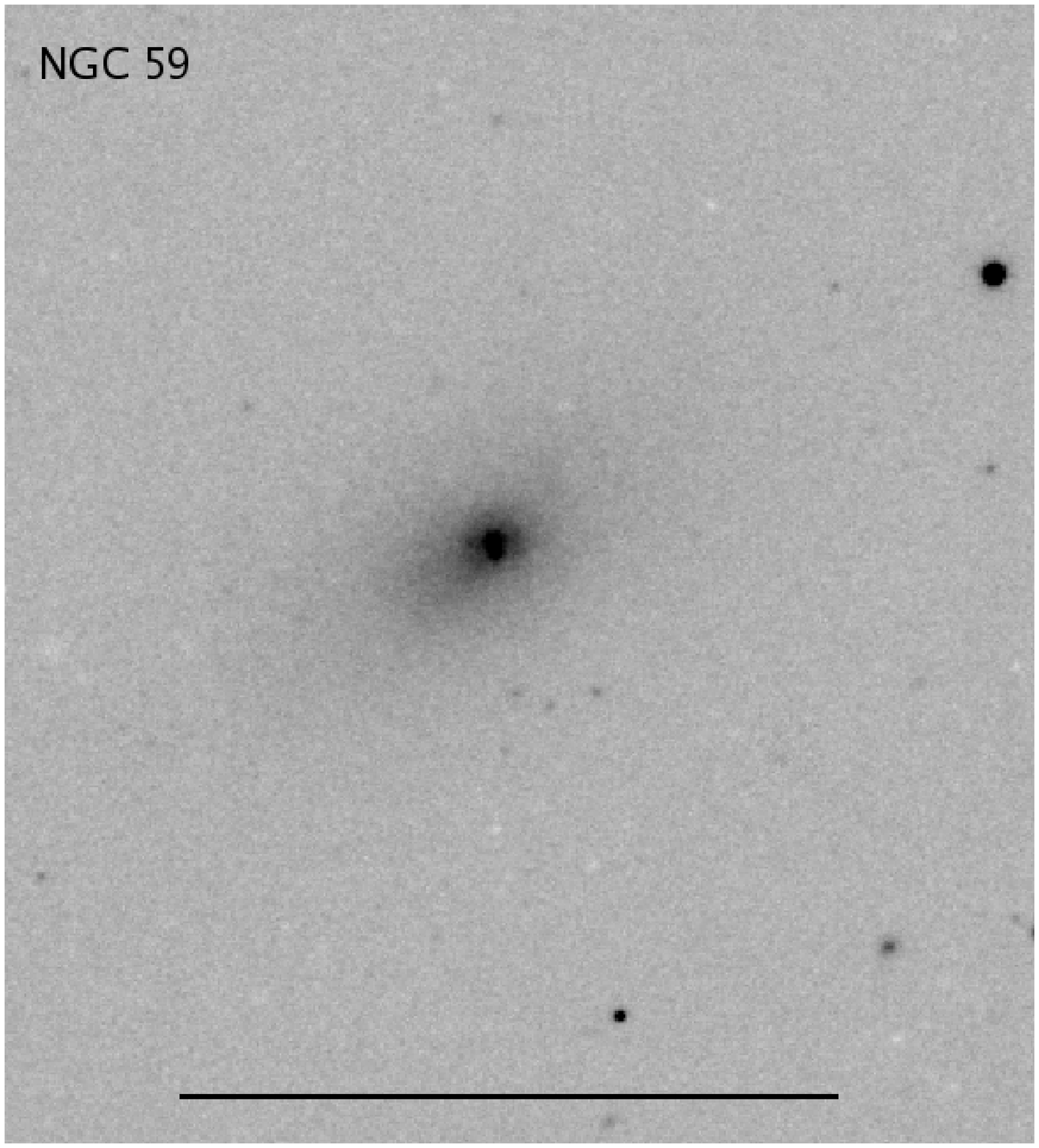}
    \includegraphics[width=7.5cm]{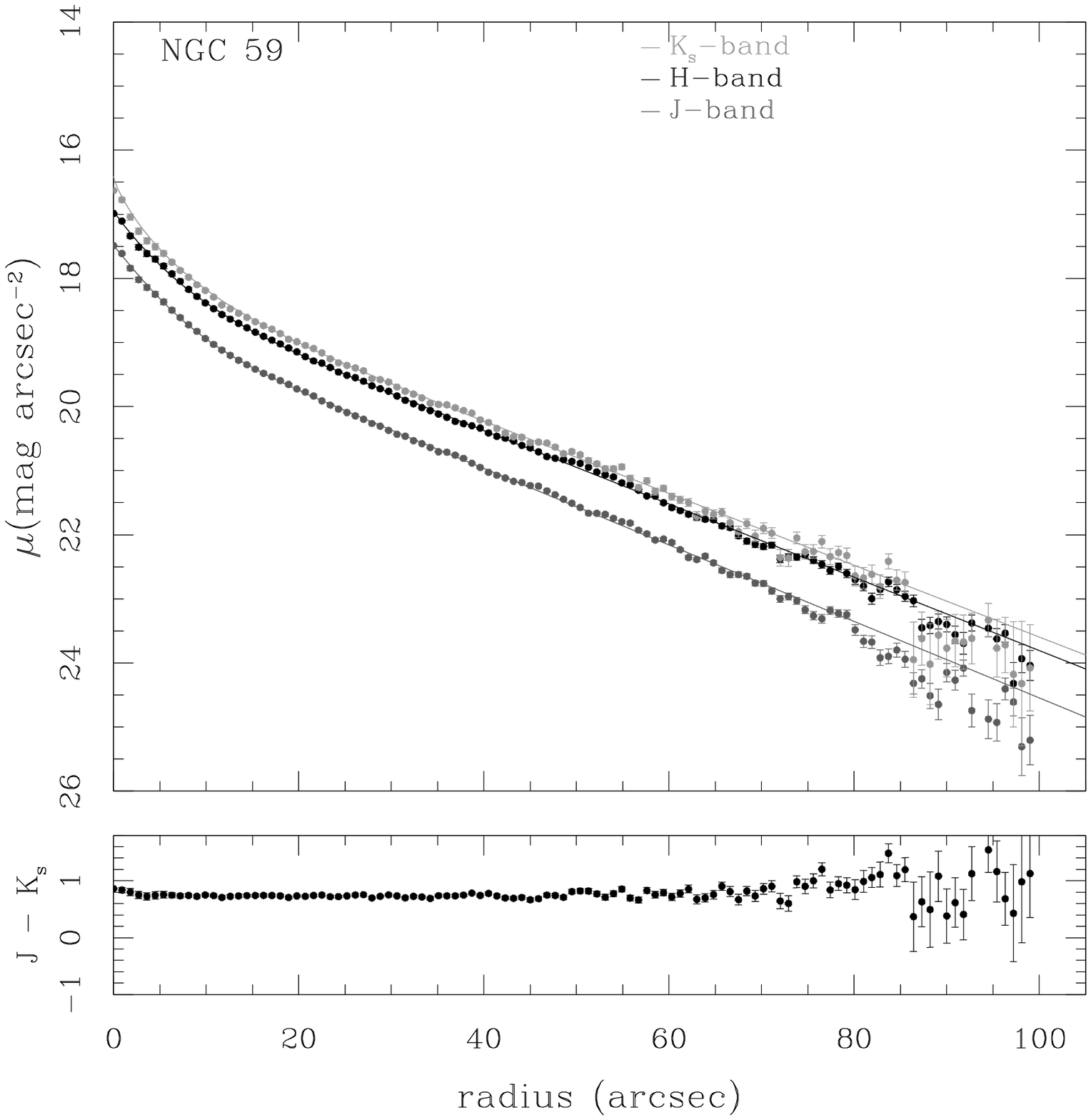}
    \includegraphics[width=6.5cm]{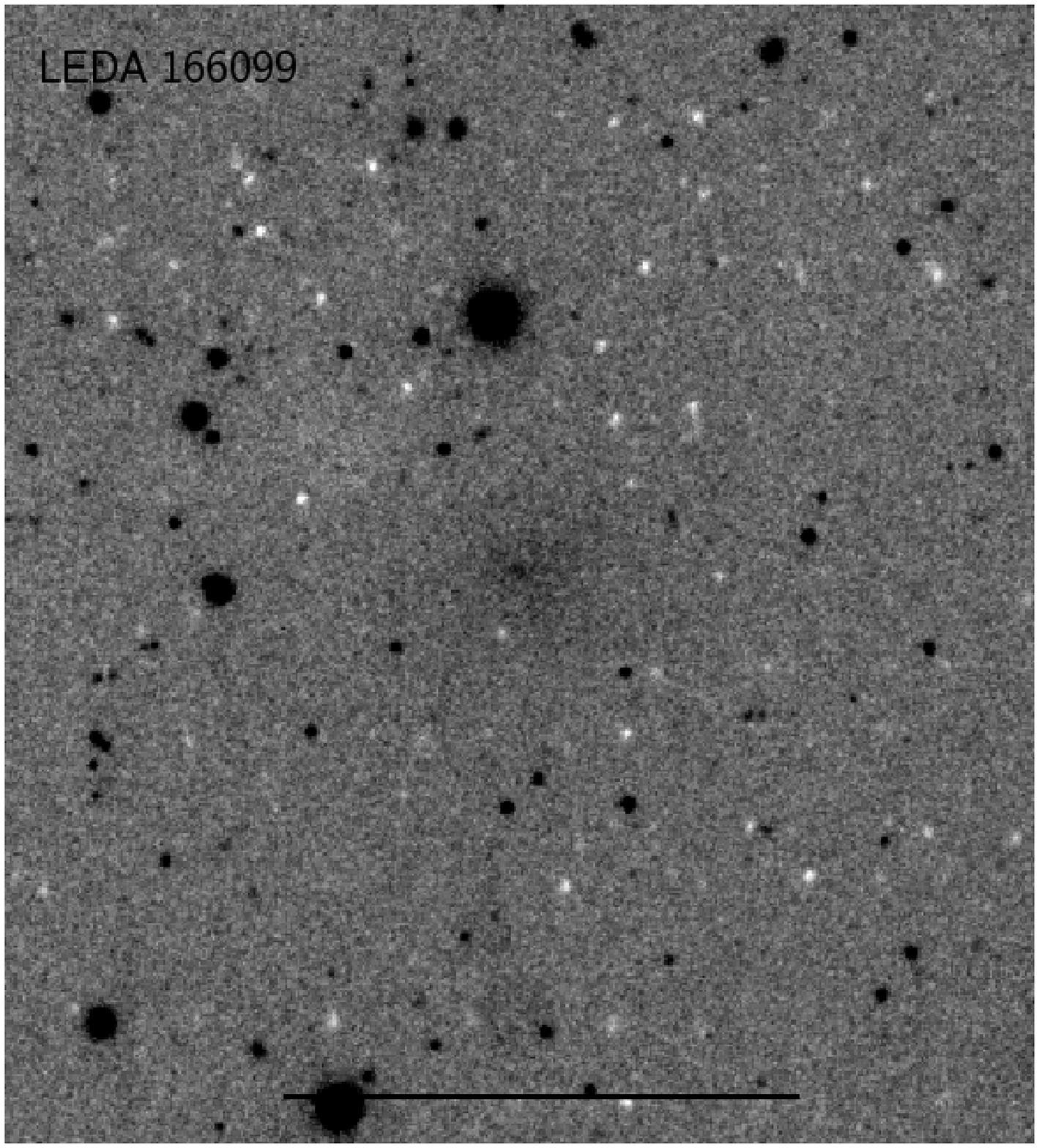}
    \includegraphics[width=7.5cm]{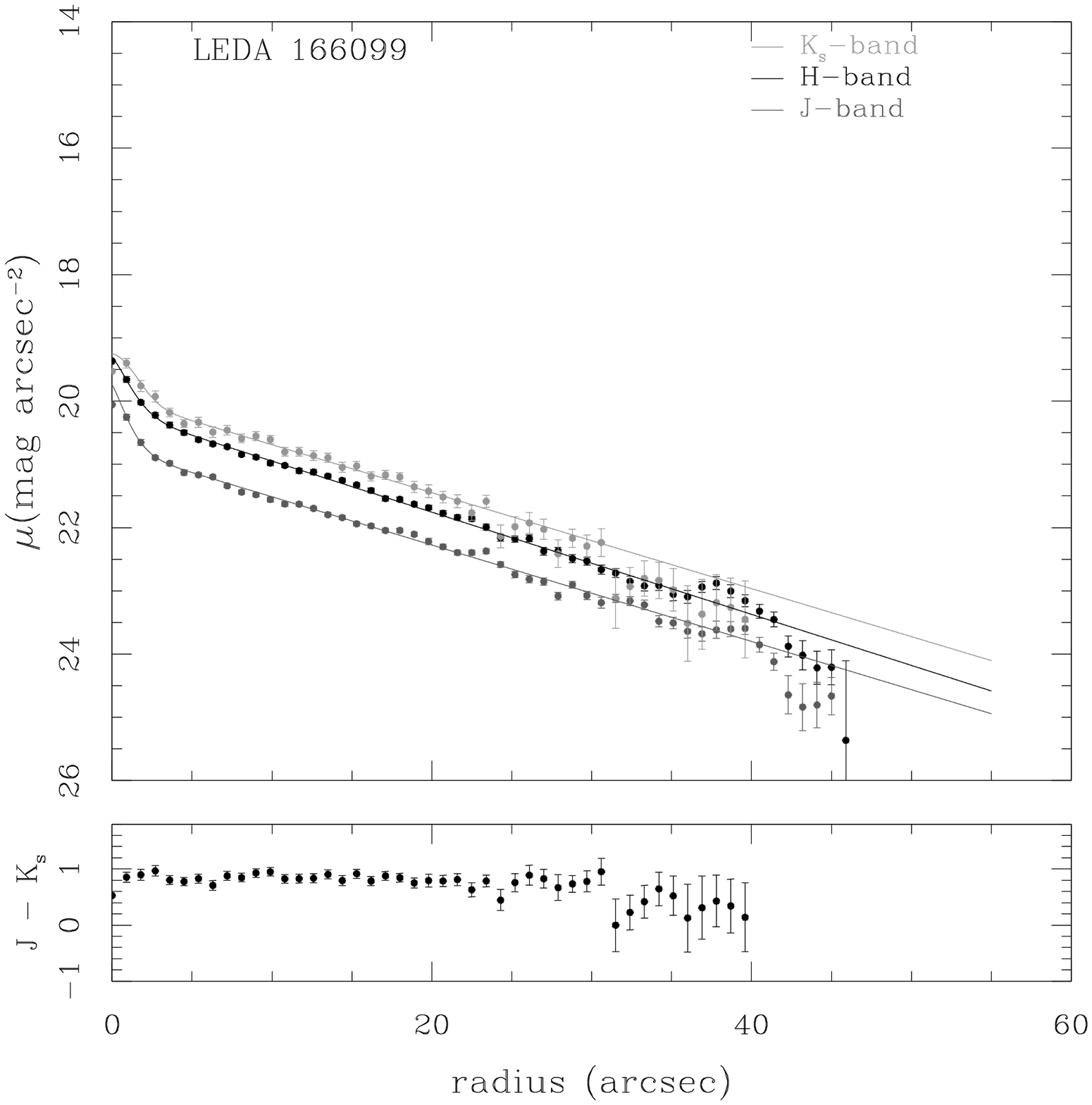}
\caption{The reduced $K_s$-band image of the dwarf galaxies is shown on the left. The horizontal bar at the bottom of each image indicates a scale of 2\arcmin. North is up and East is left. The $J$-, $H$- and $K_s$-band SB profiles of the galaxies are shown on the right. The corresponding $J$-$K_s$ colour profile is displayed below the SB profiles of the galaxy.}
\label{NIR_im1}
\end{figure*}

\begin{figure*}
    \includegraphics[width=6.5cm]{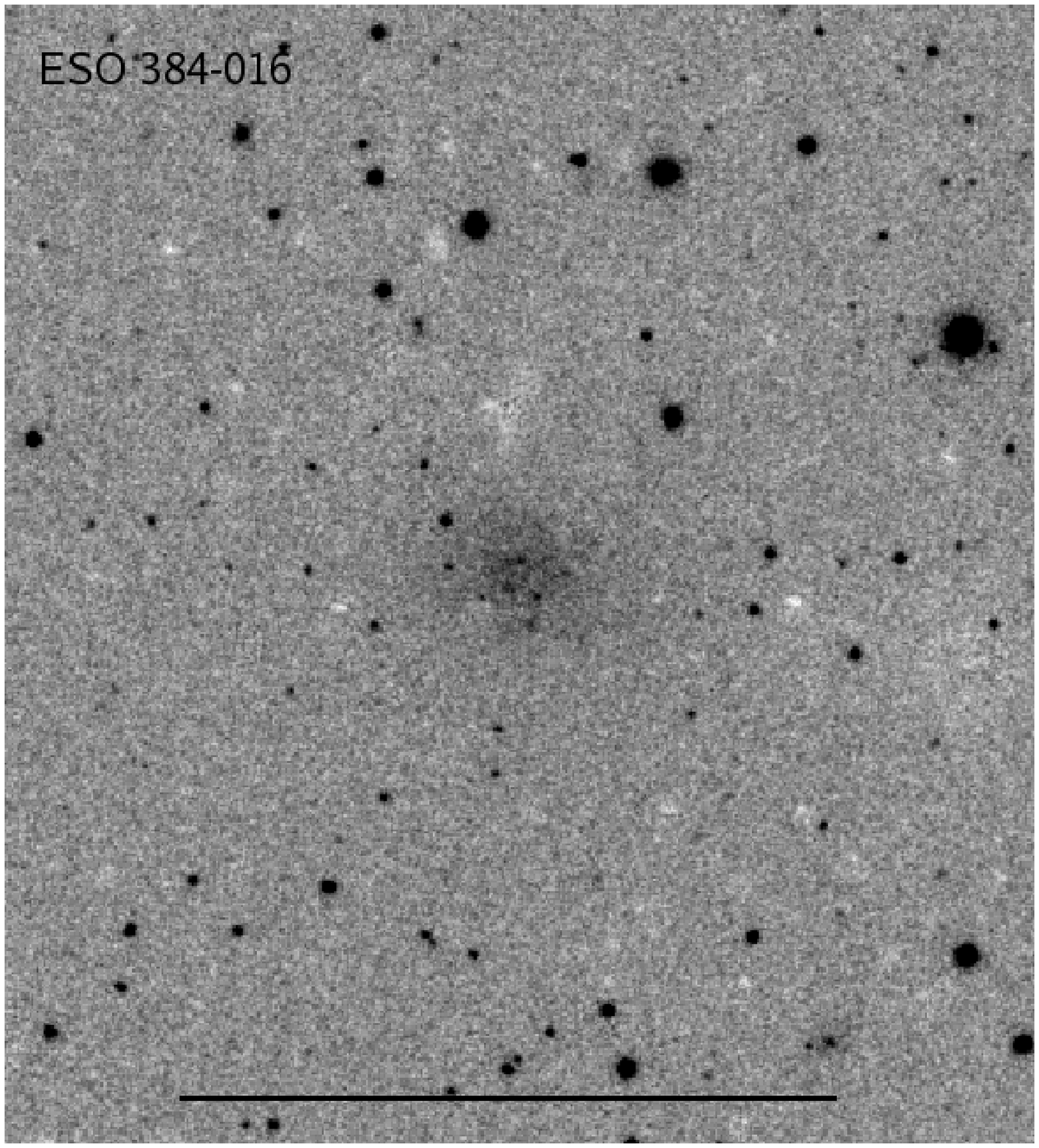}
    \includegraphics[width=7.5cm]{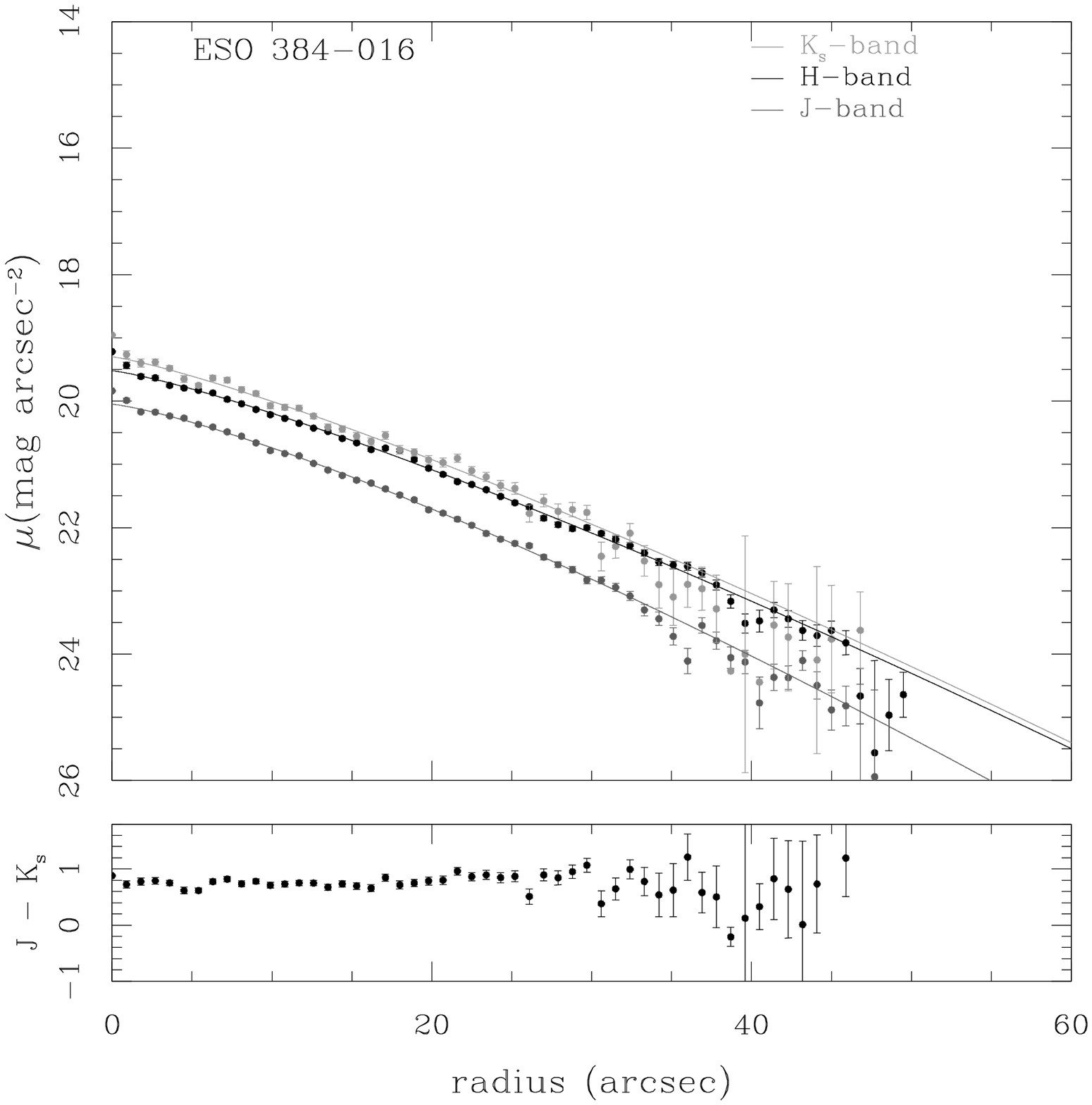}
    \includegraphics[width=6.5cm]{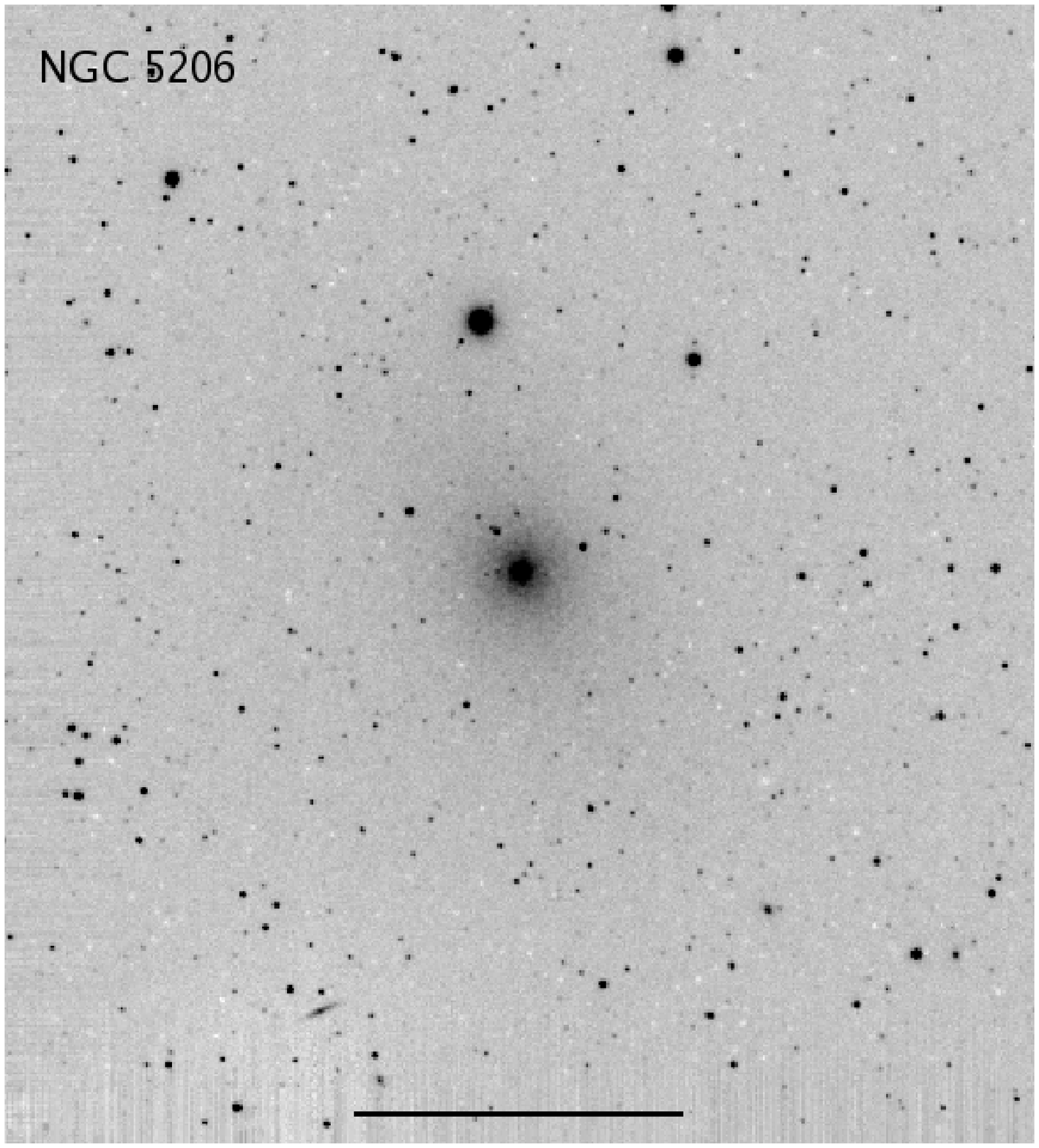}
    \includegraphics[width=7.5cm]{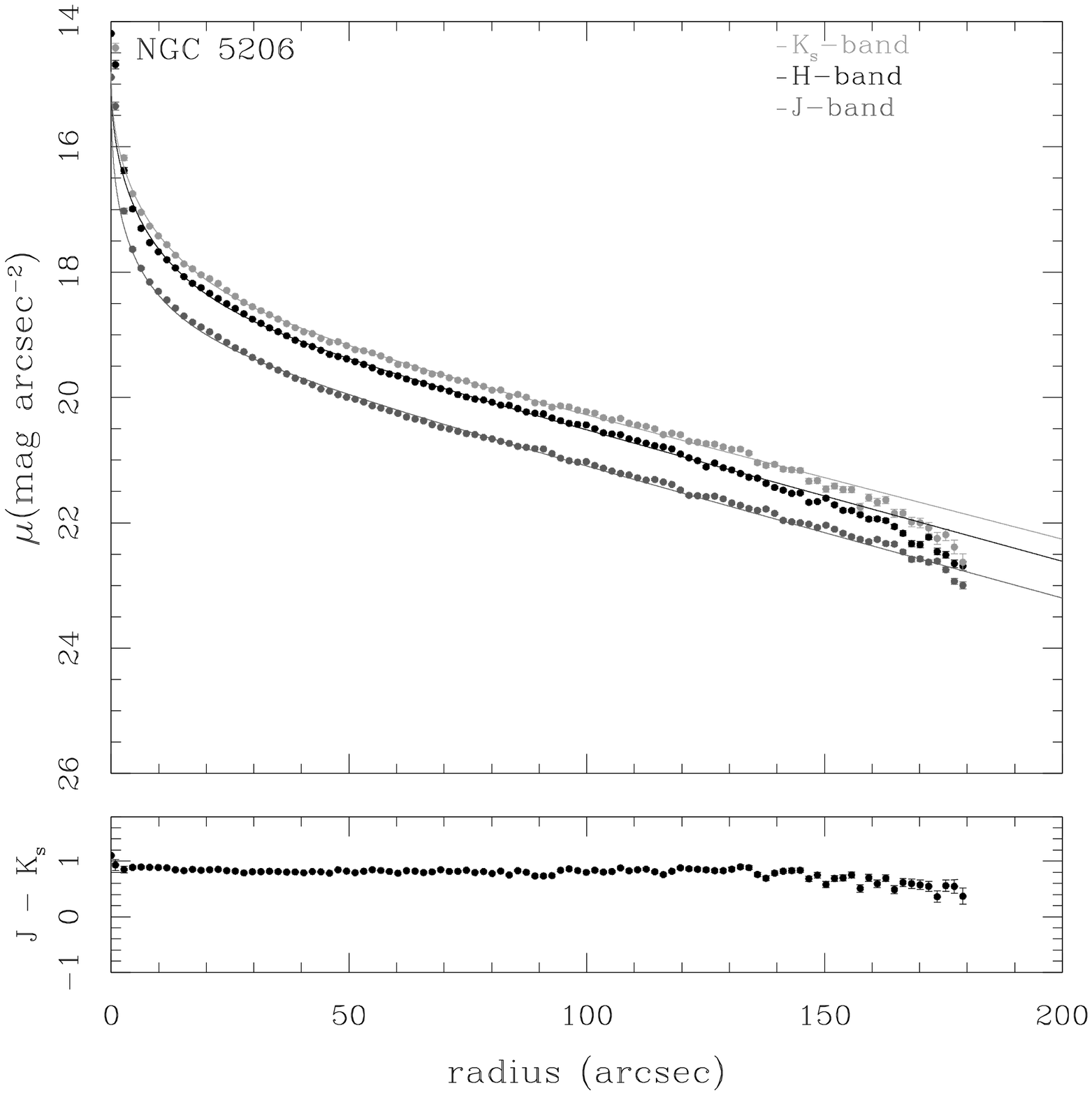}
    \includegraphics[width=6.5cm]{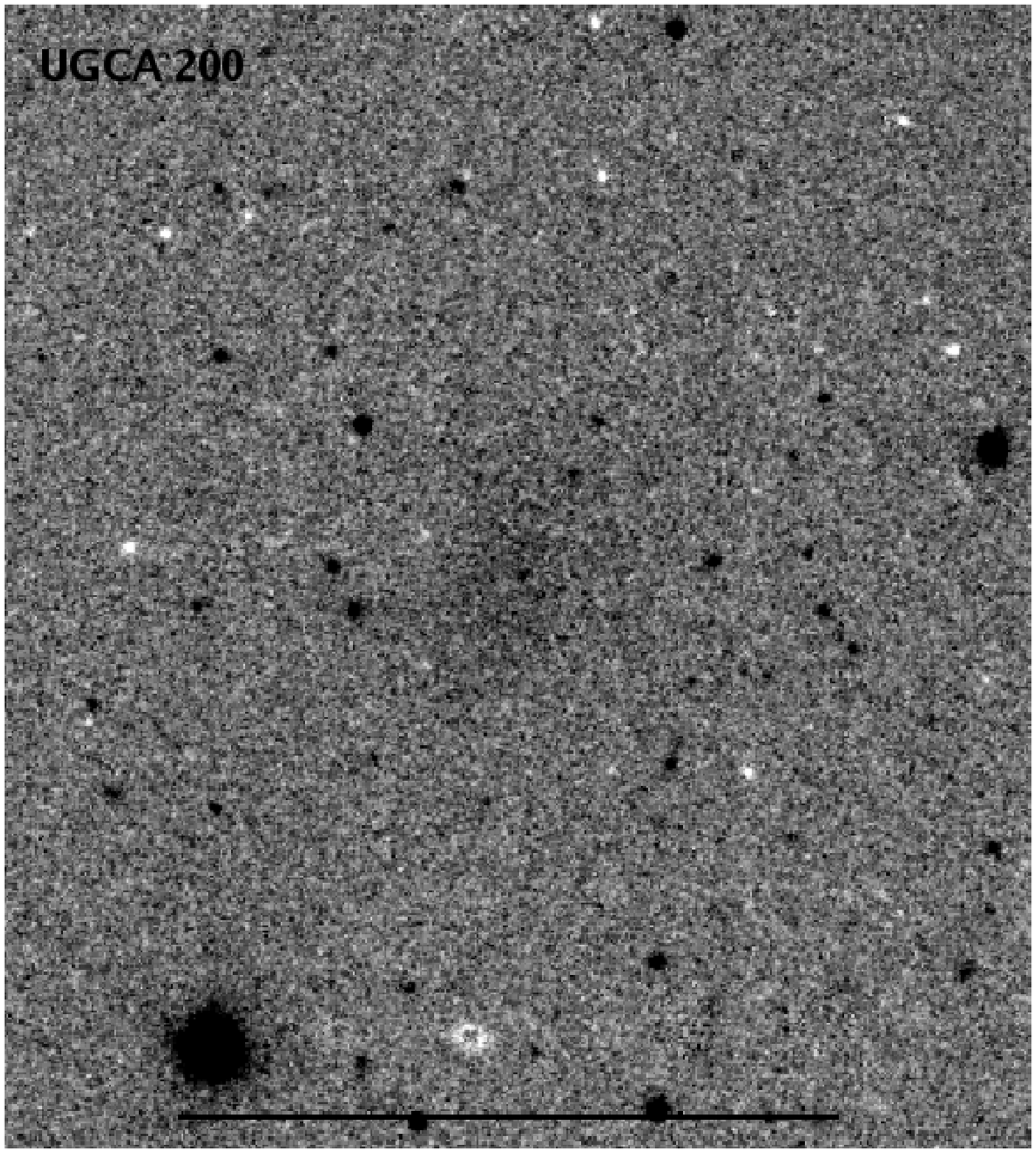}
    \includegraphics[width=7.5cm]{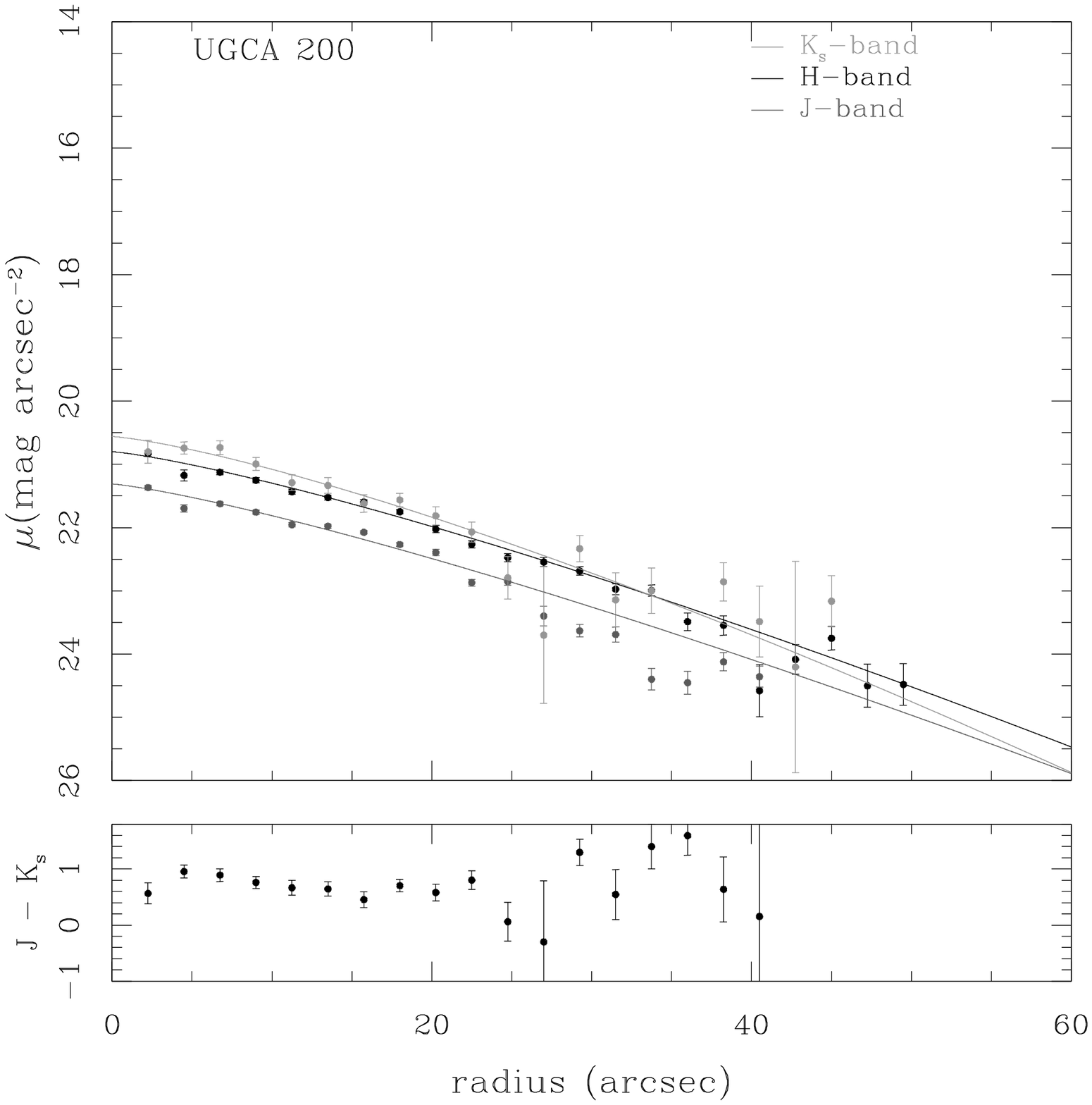}

\contcaption{}
\end{figure*}

The final term in equations (\ref{jcalib})--(\ref{kcalib}) represents the colour correction to the magnitudes. The colour coefficients for the IRSF observations have been found to be $j_2=-0.018\rm~mag$, $h_2=0.050\rm~mag$ and $k_2=0.079\rm~mag$ in the $J$-, $H$- and $K_s$-bands \citep{Kot07}, respectively. A correction for the airmass was not directly applied to the instrumental magnitudes. This calibration term is accounted for in the nightly zero-point correction derived from the 2MASS apparent magnitudes. Overall, the IRSF apparent magnitudes derived from equations (\ref{jcalib})--(\ref{kcalib}) agree within $0.05\rm~mag$ with those given by 2MASS.

\subsection[]{Star Subtraction}

\begin{figure}
    \includegraphics[width=4.1cm]{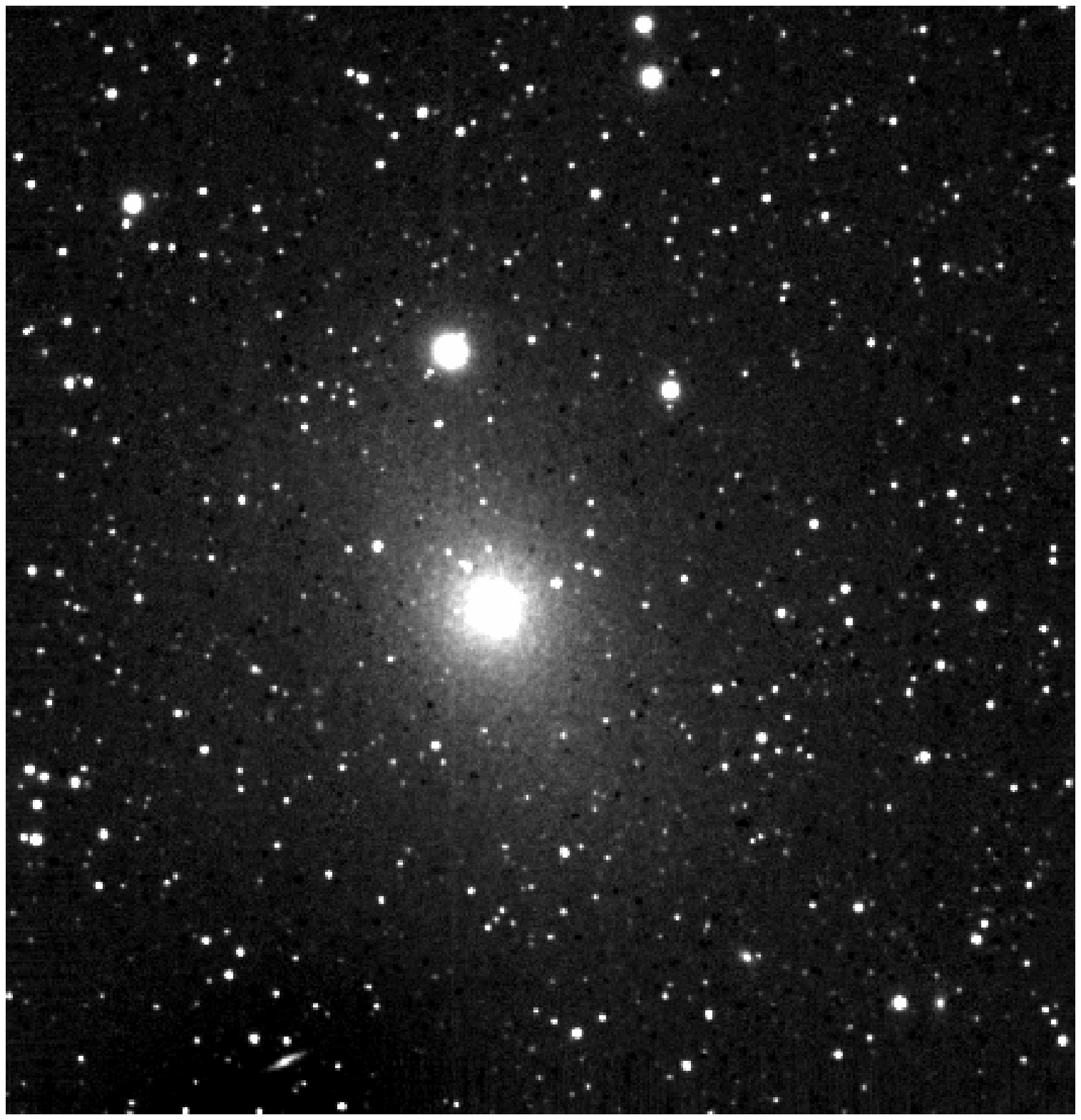}
    \includegraphics[width=4.1cm]{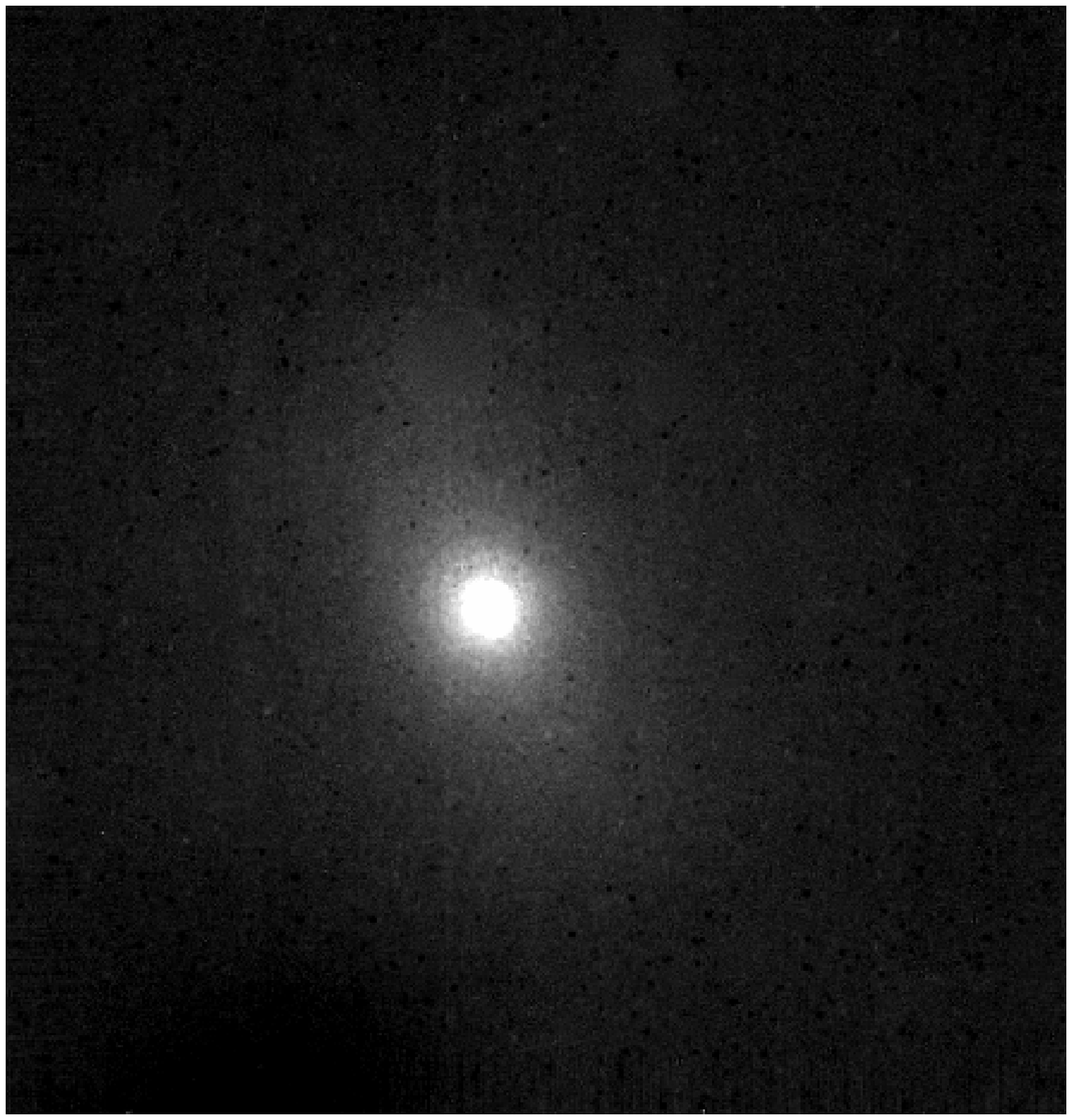}
\caption{The original (\textit{left}) and star-subtracted image (\textit{right}) of the galaxy NGC~5206. The surface photometry was performed on the cleaned galaxy image.}
\label{starsub}
\end{figure}

The $K_s$-band galaxy images in Fig.~\ref{NIR_im1} show that the foreground contamination varies from one image to the next. A careful removal of the foreground stars is essential to obtain photometric results not influenced by resolved stellar sources. The star subtraction routine of T. Nagayama (Kyoto University, private communication) was used in removing the foreground stars from the galaxy images. This routine builds the point spread function (psf) of the image by carrying out three iterative runs of the psf-fitting to the stars. This ``automated" building of the psf distinguishes the star subtraction routine of Nagayama from the conventional \textsc{killall}\ routine of \citet{But1999}. The effectiveness of the star-subtraction routine is illustrated in Fig.~\ref{starsub} which displays the original and star-subtracted image for the galaxy NGC~5206. Most of the stellar sources are cleanly removed by the routine. The residuals from background galaxies, bright and saturated stars were interactively removed using the IMEDIT task in \iraf. 

An extended source can vaguely be seen south-west of the centre of NGC~3115~DW01 (see Fig.~\ref{NIR_im1}). The NIR colour of this source indicates that it is a background galaxy which is seen through the fairly bright central region of NGC~3115~DW01. The $B$-band image of NGC~3115~DW01 from \citet{Par2002} reveals that this source is indeed a background spiral galaxy. The background galaxy was retained in the NIR images of NGC~3115~DW01 to minimize the effect of smoothing on the light distribution of the galaxy.

\section[]{Surface Photometry}  \label{sec5:SBprofiles}

The ELLIPSE\ task in \iraf\ was used to measure the intensity distribution along the semi-major axis of the galaxy. The fitting process requires a measure of the position angle of the galaxy on the sky, as well as its ellipticity. These parameters were derived from the outer isophotes where the older stellar population is expected to set the overall geometry of the galaxy. The position angle (PA) and ellipticity ($\epsilon$) of the galaxies are listed in Table~\ref{2mass_comp} which represent the mean value obtained from the three NIR bands. The $B$-band value of the position angle and ellipticity given by \citet{Par2002} is adopted for UGCA~200 whose low signal-to-noise (S/N) levels do not allow for the measurement of these parameters. By keeping the geometrical parameters fixed it is possible to measure the intensity distribution of the galaxy out to even fainter levels. The radial surface brightness of the dwarf galaxies is measured down to $\mu_{\rm lim}\simeq24\rm~mag~arcsec^{-2}$ in the $J$- and $H$-bands, and $\mu_{\rm lim}\simeq23\rm~mag~arcsec^{-2}$ in the $K_s$-band. The near-infrared SB profiles of the six dwarf galaxies are displayed in Fig.~\ref{NIR_im1}. The error associated with each point in the SB profile was computed as the RMS scatter in intensity given by the isophotal fitting.


The NIR light profiles of the galaxies NGC~3115~DW01, NGC~59, NGC~5206 and LEDA~166099 can be divided into two components: the nuclear and low SB component. The SB profiles for these galaxies show a distinct nucleus whereas pure exponential profiles are observed for ESO~384-016 and UGCA~200. The background galaxy can be identified in the $J$- and $H$-band SB profiles of NGC~3115~DW01 which show a slight increase at the location of the galaxy. NGC~5206 is the most spatially extended galaxy in the sample reaching the $\mu_J=23\rm~mag~arcsec^{-2}$ isophote at $r$$\sim$3\arcmin. The central SB component of UGCA~200 lies up to $\sim3\rm~mag~arcsec^{-2}$ above the NIR detection limit. The SB profile for this faint galaxy is measured out to a radius of $r$=40--50\arcsec\ despite its extremely low intensity levels. 

\begin{table*}
 \centering
 \begin{minipage}{170mm}
 \caption{Measured parameters for six dwarf galaxies from deep near-infared imaging: position angle (PA) measured from North to East, ellipticity ($\epsilon$), total magnitude ($m_t$) and corresponding aperture radius ($r_t$). The photometric parameters from the 2MASS All-Sky Extended Source Catalogue are listed for the brightest dwarf galaxies ($B\la13.4\rm~mag$) in the sample. The total magnitudes together with the distances from Table~\ref{dwarf_properties1} were used to compute the absolute magnitudes $M_{\rm abs}$ in the different wavelength bands. The extinction coefficients $A_\lambda$ from \citet{Sch1998} are listed. \label{2mass_comp}}
  \begin{tabular}{@{}lcccrccccccrc@{}}
\hline
	& \multicolumn{7}{c}{IRSF} & \multicolumn{5}{c}{2MASS} \\
	& {PA} &  &  & {$m_t~~~~~$} & {$r_t$} &  $M_{\rm abs}$ & $A_\lambda$  & {PA} &  &  & {$m_t~~$} & {$r_t$} \\ 
{Galaxy} & {(deg)} & {$\epsilon$} & {Filter} & {(mag)$~~~~$} & {(arcsec)}  & (mag) & (mag) & {(deg)} & {$\epsilon$} & {Filter} & {(mag)} & {(arcsec)} \\
\hline
NGC~3115~DW01 &	7 & 0.14 & $J$ & $10.70\pm0.06$ & 90 & $-19.23\pm0.20$ & 0.05  & 45 & 0.32 & $J$ & 10.99 & 70 \\
 &	& & $H$ & $10.15\pm0.07$ & & $-19.78\pm0.20$ & 0.03 & & &  $H$ & $10.15$ &  \\
 &	& & $K_s$ & $9.94\pm0.07$ & &  $-19.99\pm0.20$ & 0.02  & & & $K_s$ & $9.94$ &  \\
NGC~59 & 121 & 0.41 & $J$ & $10.84\pm0.03$ & 99 & $-17.38\pm0.16$ & 0.02 &  115 & 0.50 & $J$ & $10.89$ & 89  \\
 &	& & $H$ & $10.24\pm0.03$ & & $-17.98\pm0.16$ & 0.01 &   & & $H$ & $10.26$ &  \\
 &	& & $K_s$ & $10.07\pm0.03$ & & $-18.15\pm0.16$ & 0.01   & & & $K_s$ & $10.10$  &  \\
LEDA~166099 & 125 & 0.34 & $J$ & $13.72\pm0.03$  & 45& $-16.24\pm0.20$ & 0.18 & -- & -- &  &  &   \\
 &	& & $H$ & $13.18\pm0.04$ & & $-16.78\pm0.20$ & 0.11 &  &  & &   \\
 &	& & $K_s$ & $12.95\pm0.04$ & &  $-17.01\pm0.20$ & 0.07 &  & &  & \\
ESO~384-016 & 82 & 0.31 & $J$ & $13.12\pm0.05$ & 54& $-15.00\pm0.13$ & 0.07  & -- & -- &  &  &   \\
 &	& & $H$ & $12.49\pm0.07$ &  & $-15.63\pm0.13$ & 0.04 &  &  & &  \\
 &	& & $K_s$ & $12.35\pm0.07$ & &  $-15.77\pm0.13$ & 0.03 &  & &  & \\
NGC~5206 & 23 & 0.35 & $J$ & $8.91\pm0.04$ & 180 &  $-18.87\pm0.20$ & 0.11 & 45 & 0.16 & $J$ & 9.39 & 114  \\
 &	& & $H$ & $8.35\pm0.05$ & &  $-19.43\pm0.20$ & 0.07 &  &  & $H$ & 8.55 &  \\ 
 &	& & $K_s$ & $8.05\pm0.05$ & & $-19.73\pm0.20$ & 0.04 &   & & $K_s$ & 8.49 & \\
UGCA~200 & -31 & 0.30 & $J$ & $13.84\pm0.06$ & 45& $-16.09\pm0.20$ & 0.04  & -- & -- &  &  &   \\
 &	& & $H$ & $13.32\pm0.07$ & & $-16.61\pm0.20$ & 0.03 &  & &   \\
 &	& & $K_s$ & $13.26\pm0.09$ &  &$-16.67\pm0.20$ & 0.02  & &  &  \\
\hline
\end{tabular}
\end{minipage}
\end{table*}

The NIR colour profiles (\ie $J\mbox{-}K_s$, $H\mbox{-}K_s$ and $J\mbox{-}H$ profiles) were derived by subtracting the SB profiles as function of radius in the respective wavelength bands. We only display the $J\mbox{-}K_s$ colour profile below each of the galaxy SB profiles in Fig.~\ref{NIR_im1}. Generally, the galaxy colours remain almost constant for surface brightnesses of $\mu_{K_s}\la 22\rm~mag~arcsec^{-2}$ after which the noise levels dominate the $K_s$-band photometry. The NIR colour of the extremely low SB galaxy, UGCA~200, shows the largest scatter from the mean colour of up to $0.25\rm~mag$ within the $\mu_{K_s}\la 22\rm~mag~arcsec^{-2}$ limit. 

The mean colour of each galaxy was computed by averaging the data points above the $\mu_{K_s}\sim 22\rm~mag~arcsec^{-2}$ detection limit. The mean colours for each galaxy are listed in Table~\ref{dwarf_colors}. The six dwarfs exhibit typical NIR colours observed for dwarf galaxies in the 2MASS Extended Source Catalogue \citep{Jar2000}. The galaxies NGC~3115~DW01, LEDA~166099 and NGC~5206 are found to have redder colours of $J\mbox{-}K_s > 0.8\rm~mag$, which is in agreement with their optical $B\mbox{-}R$ colours shown in Table~\ref{dwarf_properties1}. A mean colour of $0.7 < J\mbox{-}K_s < 0.8\rm~mag$ is measured for the remaining galaxies with NGC~59 and ESO~384-016 showing similar NIR colours. 


\subsection[]{Total Magnitudes} \label{Total_mag}

The total apparent magnitude is given by the integrated flux within the detection limit of the galaxy. A growth curve was constructed for each galaxy by integrating the intensity in circular apertures. The apertures were defined at a radius step of $r=\sqrt{ab}$ where $a$ and $b$ are the respective major and minor axis of the galaxy. The total apparent magnitude, $m_t$, corresponds to the asymptotic intensity of the growth curve which is measured down to the background level of the image. A good measure of the background level was found by systematically varying the sky brightness in the image. The growth curve converges asymptotically to a flat background when the correct sky level is achieved. 

The total magnitudes ($m_t$) of the six dwarf galaxies are listed in Table~\ref{2mass_comp}. The accuracy of the total magnitudes depends on the data reduction and calibration procedures. The sky subtraction technique gives an uncertainty of $\la 0.02\rm~mag$ in the background level for the three wavelength bands. A larger source of uncertainty is introduced by the photometric calibrations of the images with the accuracy in the measured zero points varying from $0.03\rm~mag$ in the $J$-band to $0.09\rm~mag$ in the $K_s$-band. The total magnitudes are thus derived with an accuracy of $<0.1\rm~mag$ given the errors introduced through the reduction and calibration procedures.

\begin{figure}
    \includegraphics[width=4.1cm]{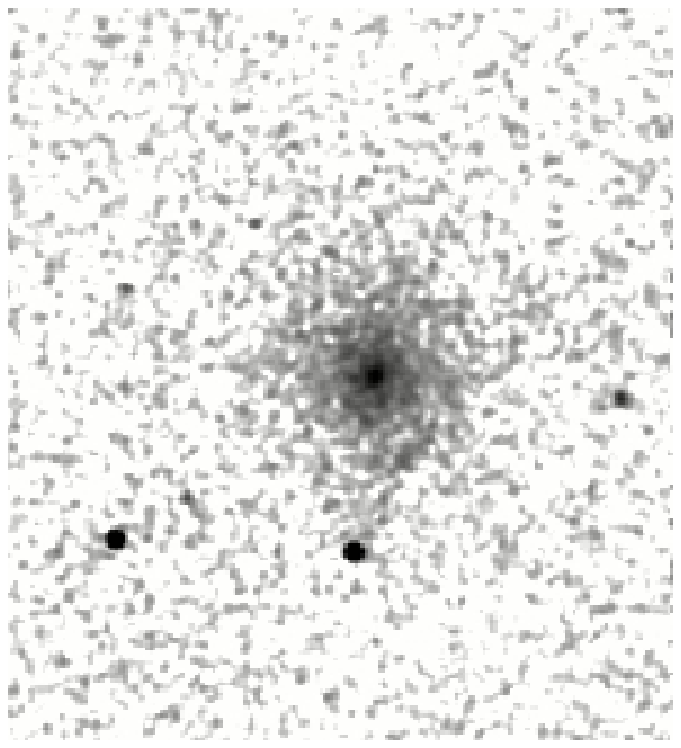}
    \includegraphics[width=4.1cm]{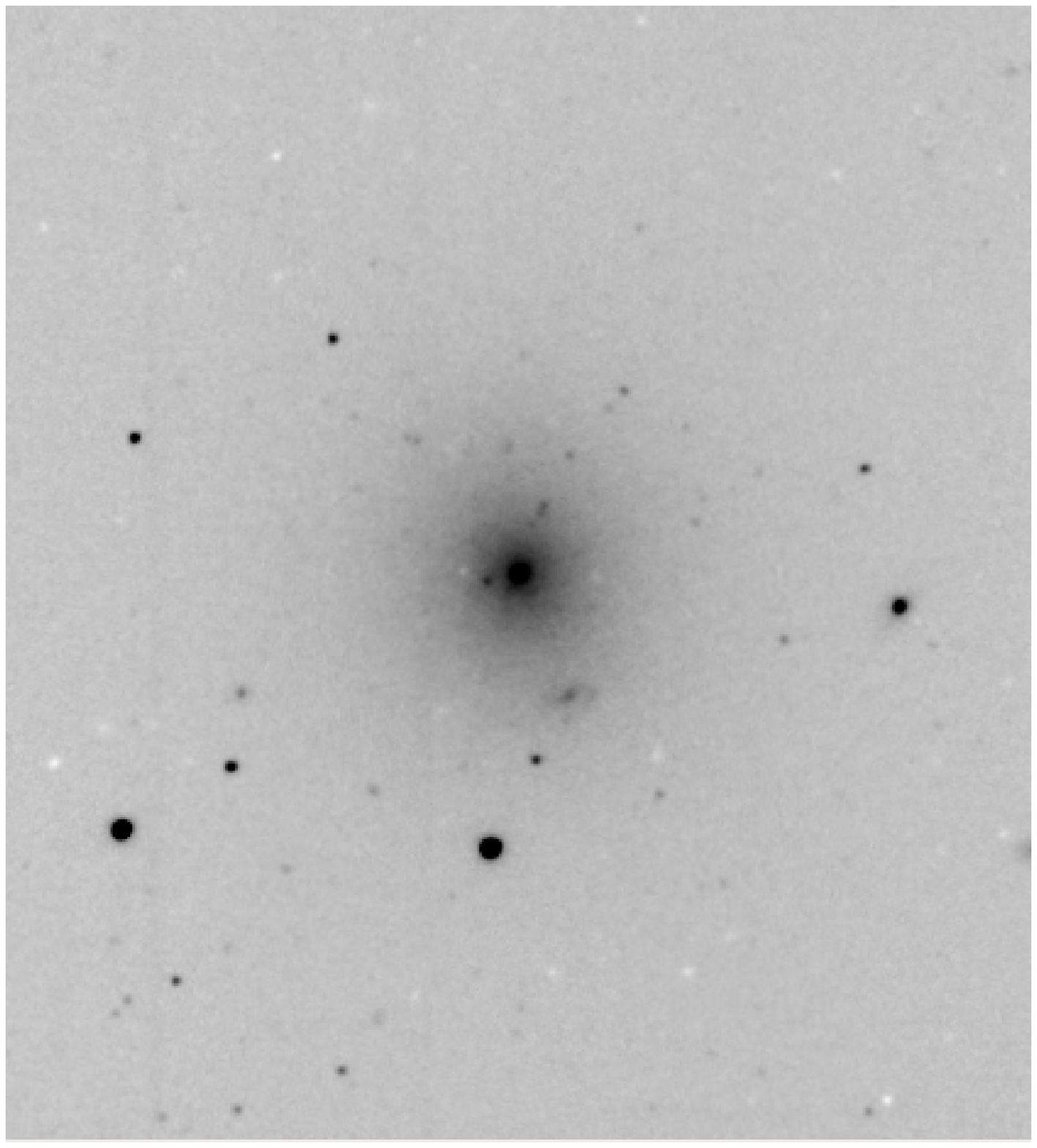}
\caption{A comparison of the $H$-band image of the galaxy NGC~3115~DW01 obtained from 2MASS (\textit{left}) and the IRSF (\textit{right}). The image size is 3\farcm5$\times$3\farcm5. North is up and East is left.}
\label{IRSFvs2MASS}
\end{figure}

NIR imaging and photometry are available for the three brightest dwarfs ($B\la13.4\rm~mag$) in our sample from the 2MASS All-Sky Extended Source Catalogue\footnote{The 2MASS Extended Source Catalogue can be accessed online at http://irsa.ipac.caltech.edu/applications/Gator/.}. A comparison of the IRSF and 2MASS $H$-band image of the galaxy NGC~3115~DW01 is shown in Fig.~\ref{IRSFvs2MASS}. The four times higher spatial resolution of the IRSF images gives a clear distinction between the nucleus and the extended low SB component of the galaxy. The 2MASS total magnitudes for the galaxies NGC~3115~DW01, NGC~59 and NGC~5206 are listed in Table~\ref{2mass_comp}. These magnitudes are the extrapolated total magnitudes $r_{\rm ext}$ in the 2MASS Extended Source catalogue and correspond to the total integrated flux of the galaxy measured down to the background level. The deep observations of the dwarf galaxies allow for their detection out to larger radii compared to 2MASS. In particular, the IRSF observations of NGC~5206 show that this galaxy has an extended low SB component which was not completely detected by 2MASS.

\begin{figure}
    \includegraphics[width=8cm]{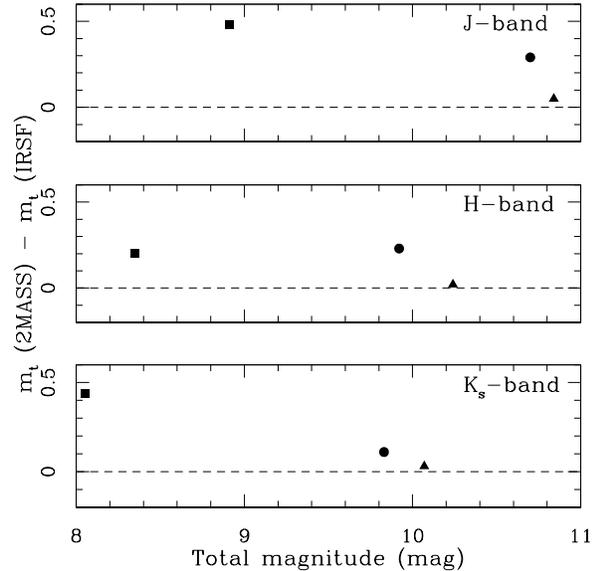}
\caption{The difference between the IRSF and 2MASS total magnitudes for the galaxies NGC~3115~DW01 (\textit{circle}), NGC~59 (\textit{triangle}) and NGC~5206 (\textit{square}) is plotted against the IRSF total magnitude.}
\label{2masscomp}
\end{figure}

The difference between the IRSF and 2MASS total magnitudes for the three dwarfs is plotted in Fig.~\ref{2masscomp}. It is seen that 2MASS underestimates the flux of the three galaxies by up to $\la0.5\rm~mag$. The largest deviation of $J=0.48\rm~mag$ from the 2MASS total magnitude is observed for the galaxy NGC~5206 which shows an extended low SB component in the deep NIR images. This underestimation of the galaxy fluxes obtained by 2MASS has been observed by \citet{Kir2008} and \citet{And2002}. Overall, the 2MASS survey fails to detect low SB galaxies so that the NIR photometric measurements of the galaxies ESO~384-016, LEDA~166099 and UGCA~200 are presented here for the first time. The deep NIR observations emphasize the serious selection biases of the 2MASS galaxy survey.

\subsection[]{Structural Parameters from One-dimensional Analytical Profile Fits} \label{sec6:SBfits}

The $J$-, $H$- and $K_s$-band SB profiles (shown in Fig.~\ref{NIR_im1}) were fit with an analytical function to characterize the underlying structure of the six dwarf galaxies. Those galaxies having a nuclear component (NGC~3115~DW01, NGC~59, LEDA~166099 and NGC~5206) were fit with a combination of an exponential and S\'ersic law to model the stellar disk component and nucleus, respectively. The S\'ersic law can take the form of \citep{Cao1993} : 
\begin{equation}
I(r) = I_e{\rm exp}\left[-b_n\left( \left(\frac{r}{r_e}\right)^{1/n} - 1\right)\right]~, \label{sersic}
\end{equation}
where $I_e$ is the intensity at the effective radius $r_e$ which contains half of the total integrated light from the model. These parameters were derived for the stellar disk of the galaxy by setting the shape parameter $n = 1$ giving an exponential profile fit. The parameter $b_n$ is directly related to the shape parameter $n$ and can be approximated by $b_n\simeq1.9992n - 0.3271$ for $0.5 \la n \la 10$. The light profiles of the remaining dwarf galaxies in the sample were appropriately modeled using a single S\'ersic function as given by equation~(\ref{sersic}).

The fits to the radial SB profiles were performed using the NFIT1D task in \iraf\ which employs a $\chi^2$-minimization technique in finding the best-fit to the observed light profiles. An exponential and S\'ersic function were fit simultaneously to the SB profiles of galaxies hosting a nucleus. For these galaxies, the stellar disk can be modeled by an exponential law out to the detection limit. The central region of the light profiles was fit with a S\'ersic function to model the steep rise in the SB profile towards the center of the galaxy. The structural parameters (\ie the effective radius $r_e$ and corresponding SB $\mu_e$) of the extended stellar disk are given in Table~\ref{1Dfits} for the nucleated dwarf galaxies. A less than $\sim1\%$ variation is observed in these parameters if instead a single exponential law is fit to the light profile. This implies that the derived structural parameters are associated purely with the stellar disk as they are not sensitive to the S\'ersic fit in the central parts of the galaxy. The structural parameters for the non-nucleated dwarfs ESO~384-016 and UGCA~200 are given by a S\'ersic fit with shape parameter in the range of $0.7<n<0.9$ (Table~\ref{1Dfits}). The derived values of $n$ overlap with those  obtained for dwarf galaxies in the $H$-band \citep[see ][]{Kir2008}.   


\begin{table}
 \centering
 \caption{Mean colours for six dwarf galaxies. \label{dwarf_colors}}
  \begin{tabular}{@{}lcccc@{}}
\hline
        & $J\mbox{-}K_s$ & $H\mbox{-}K_s$ & $J\mbox{-}H$ \\ 
 {Galaxy} & {(mag)} & {(mag)} & {(mag)} \\
\hline
NGC 3115 DW01 & $0.89\pm0.09$ & $0.27\pm0.10$ & $0.62\pm0.09$ \\
NGC 59        & $0.73\pm0.05$ & $0.16\pm0.05$ & $0.58\pm0.04$ \\
LEDA 166099   & $0.84\pm0.09$ & $0.27\pm0.10$ & $0.58\pm0.07$ \\
ESO 384-016   & $0.74\pm0.11$ & $0.17\pm0.12$ & $0.57\pm0.09$ \\
NGC 5206      & $0.82\pm0.06$ & $0.22\pm0.07$ & $0.61\pm0.06$ \\
UGCA 200      & $0.71\pm0.17$ & $0.20\pm0.17$ & $0.50\pm0.11$ \\
\hline
\end{tabular}
\end{table}

\begin{figure}
    \includegraphics[width=8.5cm]{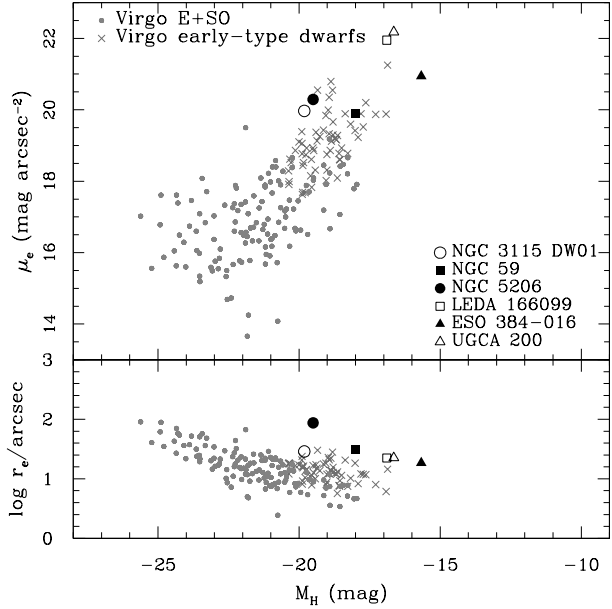}
    \caption{The effective surface brightness $\mu_e$ (\textit{top}) and effective radius $r_e$ (\textit{bottom}) as a function of $H$-band luminosity of the galaxy. The six IRSF dwarfs are shown in black where filled symbols denote those galaxies in which neutral hydrogen gas has been detected. Early-type giant galaxies (E, S0) and dwarf galaxies (dE, dS0) from the Virgo Cluster are also shown. }
\label{Hstructure}
\end{figure}

The structural parameters ($\mu_e$ and $r_e$) of the six dwarf galaxies are plotted as a function of their corresponding $H$-band luminosity in Fig.~\ref{Hstructure}. The $H$-band parameters were used for the plots as these can be directly compared to the structural parameters for early-type galaxies found in the literature. In particular, \citet{Gav2000} have carried out $H$-band surface photometry for a large sample of early-type galaxies in the Virgo Cluster. The photometric data for the galaxies can be accessed online using the Goldmine database \citep{Gav2003}.\footnote{http://goldmine.mib.infn.it/ .} The structural parameters for elliptical (E) and lenticular (S0) galaxies, as well as early-type dwarf systems (dE, dS0) from the Virgo Cluster have been added to Fig.~\ref{Hstructure}. For these galaxies, $r_e$ is defined as the radius enclosing half of the light along the observed SB profile where the total amount of light is given by the total $H$-band magnitude of the galaxy. The effective surface brightness $\mu_e$ corresponds to the mean SB within the effective radius. A mean distance of 15.8~Mpc \citep{Jer2004} was adopted for the Virgo Cluster. 

In Fig.~\ref{Hstructure} it can be seen that the six dwarfs follow the correlation between the effective SB and $H$-band luminosity observed for early-type galaxies. The IRSF dwarf galaxies fit well with the sequence of early-type dwarfs with the fainter galaxies ($M_H\ga-18\rm~mag$) in the sample forming the lower luminosity extension of the observed correlation. The bottom panel of Fig.~\ref{Hstructure} shows that the dwarf galaxies do not follow the same relation between the effective radius and $H$-band luminosity seen for the giant galaxies. This is in agreement with \citet{Zib2002} who observed a dichotomy in the $\mu_e$-$r_e$ relation for giant and dwarf galaxies in cluster environments. The IRSF dwarfs provide even further evidence of a disparate group of dwarf galaxies with effective radii comparable to some of the brighter giants in the Virgo Cluster.  

\begin{table*}
 \centering
 \caption{NIR structural parameters for the six dwarf galaxies from 1D analytical fits. Columns: (1) Galaxy designation; (2) Presence of a nucleus; (3) NIR filter/band; (4) Analytical function fit to the light profile; (5) Effective surface brightness and (6) Effective radius from the model; (7) Shape parameter: $n=1$ for an exponential fit to the disk component of the galaxy. \label{1Dfits}}
  \begin{tabular}{@{}lcccccc@{}}
\hline
	         & 	             &             &        &       {$\mu_{\rm e}$}                &   $r_{\rm e}$   &       \\
{Galaxy}  & {Nucleus} & {Filter} & {Fit} & {($\rm mag~arcsec^{-2}$)}  & {(arcsec)}     & n      \\
(1)	        & (2)	       & (3)	 & (4)     	        & (5)		    			      & 		(6)	    & 	(7)	\\
\hline
NGC~3115~DW01 & Yes & $J$     	   & Exp & 20.83 & 31.73 & 1.00     \\
	     			  &	     & $H$     & Exp  & 19.97 & 28.79 & 1.00    \\
 	     			  &	     & $K_s$ & Exp  & 19.57 & 26.52 & 1.00     \\
NGC~59 		  	   & Yes & $J$      & Exp & 20.39 & 30.30 & 1.00  \\
	     			  &	     & $H$     & Exp & 19.90 & 31.62 & 1.00   \\
 	     			  &	     & $K_s$ & Exp & 19.81 & 32.30 & 1.00   \\
LEDA~166099	  & Yes 	     & $J$      & Exp  & 22.56 & 23.71 & 1.00   \\
	     			  &	     & $H$     & Exp  & 21.95 & 22.38 & 1.00  \\
 	     			  &	     & $K_s$ & Exp  & 21.74 & 23.83 & 1.00  \\
ESO~384-016	  	  & No   & $J$      &  S\'ersic &  21.40 & 17.06 & 0.79  \\
	     			  &	     & $H$     &  S\'ersic &  20.94 & 18.45 & 0.82  \\
 	     			  &	     & $K_s$ &  S\'ersic &  20.74 & 18.10 & 0.83  \\
NGC~5206 		  & Yes & $J$     &    Exp  & 20.91 & 86.58 & 1.00   \\
	     			  &	     & $H$     &  Exp  & 20.29 & 87.26 & 1.00   \\
 	     			  &	     & $K_s$ &   Exp & 20.20 & 92.99 & 1.00   \\
UGCA~200 		  & No & $J$      &  S\'ersic    &	 22.71	& 22.93	& 0.81   \\
	     			  &	     & $H$     &  S\'ersic  &	 22.18	& 22.57	& 0.80  \\
 	     			  &	     & $K_s$ &  S\'ersic  & 21.87	& 20.42	& 0.77 \\
\hline
\end{tabular} 
\end{table*}

\section[]{Discussion} \label{sec7:NIRdiscussion}

\subsection{$J$-$K_s$ Colour-Luminosity Relation}

\begin{figure}
    \begin{center} 
    \includegraphics[width=8.5cm]{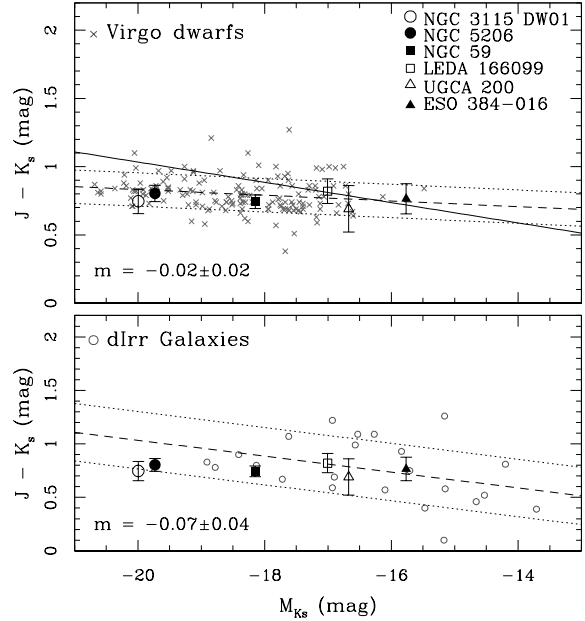}
    \caption[$J$-$K_s$ colour-magnitude diagram for early-type galaxies and dIrr's.]{$J$-$K_s$ colour-magnitude diagram for the Virgo early-type (dE, dS0 and dE/dS0) galaxies (\textit{top}) and dIrr's (\textit{bottom}). The six IRSF dwarfs are shown in both panels where the filled symbols indicate those dwarfs which have been detected in \ion{H}{i}. The linear least-squares fit and $1\sigma$ standard deviation for each galaxy sample are represented by the dashed and dotted lines, respectively. The slope $m$ of the linear fit is indicated. For comparison, the linear fit for the dIrr galaxies is re-plotted (solid line) in the top panel.}
    	\label{JK_dwarfs}
	\end{center}
\end{figure}

The $J$-$K_s$ colour-magnitude diagram for the six IRSF dwarfs is shown in Fig.~\ref{JK_dwarfs}. The mean $J$-$K_s$ colours for the galaxies (listed in Table~\ref{dwarf_colors}) are plotted against their corresponding absolute $K_s$-band magnitudes. The $k$-correction for galaxy redshifts was not applied to the magnitudes. At redshifts of $z<0.1$, the $k$-corrections are smaller than the photometric errors and can therefore be neglected. In the top panel of Fig.~\ref{JK_dwarfs}, the $J$-$K_s$ colours for the IRSF dwarfs are compared to those obtained for early-type dwarf galaxies in the Virgo Cluster (taken from the Goldmine database \citep{Gav2003}). 

Variations in the NIR colour of a galaxy are mainly driven by the metallicity of its stellar population. In particular, the $J$-$K_s$ colour index is very metal sensitive with little dependence on the mean age of the stellar population. To identify a possible trend in the $J$-$K_s$ colour of the dwarf galaxies with luminosity, a linear least squares fit was made to the Virgo dwarfs. The linear fit has a shallow slope of $m = -0.02\pm0.02$ suggesting that early-type dwarf galaxies show almost constant $J$-$K_s$ colours independent of their luminosity. The six IRSF dwarfs are found to lie well within the $1\sigma$ deviation from the linear fit so that these nearby systems follow the linear relation observed for the Virgo dwarfs. \citet{Gal2002} measure redder $J$-$K_s$ colours with increasing luminosity for a large sample of low surface brightness galaxies. They conclude that the more massive galaxies tend to have more metal-rich stellar populations. This correlation between the metallicity and galaxy mass is not, however, observed for early-type dwarf galaxies. 
 
Neutral hydrogen gas has been detected in the galaxies NGC~5206, NGC~59 and ESO~384-016 in our sample. For this reason, the $J$-$K_s$ colours of the IRSF dwarfs were compared to those obtained for gas-rich dwarf irregular (dIrr) galaxies \citep{Vad05} which is shown in the lower panel of Fig.~\ref{JK_dwarfs}. A least squares fitting was made to the dIrr galaxies which gives a slope of $m = -0.07\pm0.04$. The linear fit indicates that the dIrr galaxies show redder $J$-$K_s$ colours with increasing luminosity. The IRSF dwarfs are seen to lie within the $1\sigma$ deviation from the linear fit to the dIrr's. The redder $J$-$K_s$ colours with increasing luminosity are however not observed for the IRSF dwarfs who show more constant $J$-$K_s$ colours down to an absolute magnitude of $M_{K_s}=-15.8\rm~mag$. The correlation of the $J$-$K_s$ colour with absolute magnitude $M_{K_s}$ remains questionable for the low luminosity dIrr's ($M_{K_s}>-15.5\rm~mag$). It is not clear if a selection effect results in the lack of low luminosity, red dIrr galaxies at the faint end of this relation.

\subsection{Dwarf Galaxy Evolutionary Sequence and Morphology} 

\begin{figure}
    \begin{center} 
    \includegraphics[width=8.5cm]{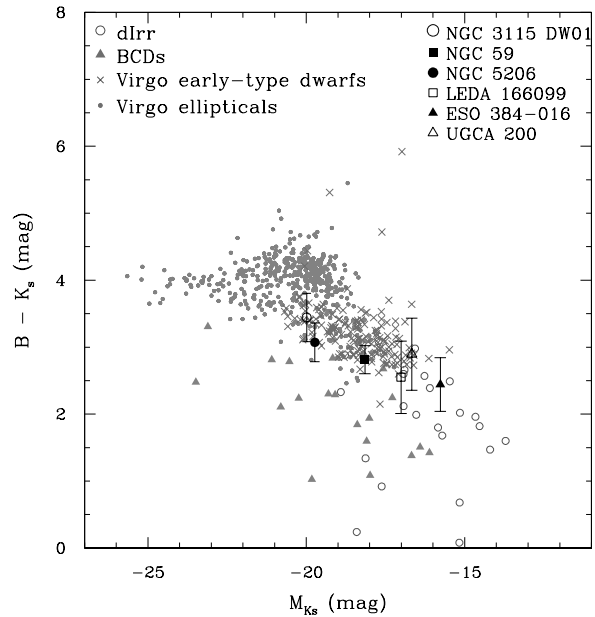}
    \caption{$B$-$K_s$ colour-magnitude diagram. The six IRSF dwarfs are shown by the black points where the filled symbols indicate those dwarfs which have been detected in \ion{H}{i}. The elliptical and early-type dwarf galaxies from the Virgo Cluster are also shown. The dIrr's \citep{Vad05} are indicated by the open circles while the BCD galaxies from \citet{Cai03} and \citet{Noe2003} are represented by the filled triangular points.}
    	\label{BK}
	\end{center}
\end{figure}

The NIR photometric results for the six dwarf galaxies were combined with existing optical measurements to assess their evolutionary state and morphology. The $B$-$K_s$ colour is known to be a good indicator of the galaxy morphological type \citep[\eg][]{Jar2003}. In Fig.~\ref{BK}, the $B$-$K_s$ colours of the six IRSF dwarfs are plotted against their corresponding $K_s$-band luminosity. The total apparent $B$-band magnitudes of the dwarfs (see Table~\ref{dwarf_properties1}) are taken from \citet{Kara2004}. For the faintest galaxies in the sample ($B\ga15.2\rm~mag$), the total $B$-band magnitudes have been determined with an accuracy of $\sim0.5\rm~mag$. The $B$-$K_s$ colours were calculated as the difference between the total apparent $B$ and $K_s$-band (from Table~\ref{2mass_comp}) magnitudes of the galaxies. 

To gain more perspective of where the IRSF dwarfs are located relative to other galaxy morphologies, we have added four different galaxy samples to the $B$-$K_s$ colour-magnitude diagram: early-type dwarfs (dE, dS0, dE/dS0) and elliptical galaxies from the Virgo Cluster (taken from the the Goldmine database), dIrr galaxies from \citet{Vad05}, blue compact dwarf (BCD) galaxies from \citet{Cai03} and \cite{Noe2003}. The general trend shows redder $B$-$K_s$ colours for the elliptical galaxies, while bluer colours are measured for both early and late-type dwarfs. An average $B$-$K_s$ colour of $\sim4\rm~mag$ was measured for early-type (E, S0) galaxies in the 2MASS Extended Source Catalog \citep{Jar2003}. The early-type dwarfs form a continuous sequence between the more luminous elliptical galaxies and late-type dwarfs. The emission from the BCD galaxies is dominated by the younger, starburst component indicated by the blue colours. Figure~\ref{BK} shows that the six IRSF dwarfs fit very well with the sequence of early-type dwarf galaxies. 

A detailed study of the relationship between the $B$- and $H$-band luminosity of galaxies was made by \citet{Kir2008}. They found a tight correlation between the absolute $B$- and $H$-band magnitudes for their sample of nearby galaxies ($D\la10\rm Mpc$). This correlation was extended to include more luminous galaxies such as the sample of spirals from \citet{Kas2006} and galaxies from the Virgo Cluster \citep{Gav2003}. The six IRSF dwarfs are found to closely follow the linear relation of \citet{Kir2008}. The correlation between the $B$- and $H$-band luminosity implies that the early-type dwarf galaxies are minimally affected by dust. This result is supported by the optical and NIR images of the IRSF dwarfs which reveal similar galaxy morphologies in the different wavelength bands. 


\subsection{Stellar Masses of the Dwarf Galaxies} \label{NIRmass}

We have adopted the galaxy evolutionary models of \citet{Bel2001} in determining the stellar $M/L$ ratio of the six dwarf galaxies. The linear coefficients given in Table~1 of \citet{Bel2001} were used to derive the $M/L$ ratios. These coefficients were measured by adopting a formation epoch model (with bursts) and a Salpeter initial mass function (IMF). It is expected that the choice of the IMF will not introduce uncertainties greater than 10\% in the NIR stellar $M/L$ ratio \citep[see][]{Gal2002}. 

\begin{table*}
 \centering
 \begin{minipage}{106mm}
 \caption{Total stellar masses ($M_{stars}$) of six IRSF dwarfs. The columns represent, (2): Extinction-corrected, absolute $H$-band magnitudes; (5): \ion{H}{i} masses of galaxies taken from the literature: NGC~59 and ESO~384-016 from \citet{Bea2006}; NGC~5206 from \citet{Cot97}. The upper \ion{H}{i} mass limit is quoted for the galaxy NGC~5206. \label{ML_ratio1}}
  \begin{tabular}{@{}lcrlrr@{}}
\hline
	& {$M_H^0$} & {$L_H$} & {$M_{stars}$} & {$M_{\ion{H}{i}}$}  &   \\
{Galaxy} & {(mag)} & {($\times10^9L_\odot$)} & {($\times10^9M_\odot$)} & {($M_\odot$)} & {$M_{\ion{H}{i}}/M_{stars}$} \\
{(1)} & {(2)} & {(3)} & {(4)} & {(5)} & {(6)} \\
\hline
NGC~3115 DW01 & -19.81 & $1.84\pm0.76$ & $2.58\pm1.17$ & -- &\\
NGC~59        & -17.99 & $0.34\pm0.13$ & $0.48\pm0.21$ & $1.5\times10^7$ & 0.031 \\
LEDA~166099   & -16.89 & $0.12\pm0.05$ & $0.17\pm0.08$ & -- & \\ 
ESO~384-016   & -15.67 & $0.04\pm0.01$ & $0.06\pm0.03$ & $6.0\times10^6$ & 0.100 \\
NGC~5206      & -19.50 & $1.38\pm0.57$ & $1.93\pm0.87$ & $5.5\times10^5$ & $<0.001$ \\
UGCA~200      & -16.64 & $0.10\pm0.04$ & $0.14\pm0.06$ & -- & \\
\hline
\end{tabular}
\end{minipage}
\end{table*}

The $H$-band stellar $M/L$ ratio of the galaxies was computed using the linear relation: 
\begin{equation} \Upsilon_H = a_H + b_H(B-H)~. \label{ML_eqn} \end{equation}
The stellar $M/L$ ratio was calculated for each of the IRSF dwarfs by substituting their individual $B$-$H$ colour into equation (\ref{ML_eqn}). To increase the statistics by using a larger galaxy sample, galaxies having absolute magnitudes in the range of $-15.6\la M_H \la -19.8~\rm mag$ in the \citet{Kir2008} sample were also used in calculating the $M/L$ ratio. These magnitudes span the range of $H$-band absolute magnitudes observed for the IRSF dwarf galaxies (see Table~\ref{2mass_comp}). A total of 33 dwarf galaxies were identified in this magnitude range.  It should be noted that the dIrr galaxy AM~0521-343 in the \citet{Kir2008} sample was excluded as the photometry of this galaxy is compromised by a bright foreground star. The $M/L$ ratio of the dwarf galaxies was computed by taking the mean of the individual ratios obtained for both the IRSF and \citet{Kir2008} galaxy samples. This gives an $H$-band stellar $M/L$ ratio of $\Upsilon_H = 1.4\pm0.8$ where the error in the $M/L$ ratio represents the standard deviation of the mean. This value overlaps with the $M/L$ ratio of $\Upsilon_H = 0.9\pm0.6$ obtained by \citet{Kir2008} for their full sample of 57 galaxies (consisting mostly of irregulars). 

The $M/L$ ratio of $\Upsilon_H = 1.4\pm0.8$ was used to calculate the stellar mass of each of the IRSF dwarfs. The extinction-corrected, absolute $H$-band magnitudes $M_H^0$ of the galaxies were converted into luminosity using the standard relation
\begin{equation} L_H = 10^{0.4(M_{H,\odot} - M_H^0)} \end{equation}
where $M_{H,\odot} = 3.35\rm~mag$ is the $H$-band luminosity of the sun \citep{Col1996}. The total stellar masses of the galaxies are listed in Table~\ref{ML_ratio1}. The two brightest dwarfs (NGC~3115~DW01 and NGC~5206) in the sample ($M_H^0<-19\rm~mag$), have the largest stellar masses which are of the order of $10^{9} M_\odot$. \citet{Puz2000} measure $(4.8\pm2.3)\times10^{10}\Msun$ as the lower mass estimate of NGC~3115~DW01. This mass estimate was obtained from the kinematics of seven GCs in the galaxy. We have derived an $H$-band stellar mass of $(2.6\pm1.2)\times10^{9}\Msun$ for NGC~3115~DW01 which is $\sim25$ times lower than the mass estimate imposed by the kinematics of the seven GCs. The dynamical mass of NGC~3115~DW01 is measured out to a projected radius of $r\simeq2.7\arcmin$, making it almost twice as large as the radius in which the stellar mass of the galaxy was determined. This result suggests that NGC~3115~DW01 is a very DM dominated galaxy with no direct evidence of an extended DM halo \citep[\eg][]{Kle2002}. On the other hand, it is also possible that the GC system of \citet{Puz2000} is not virialised around NGC~3115~DW01 giving the higher dynamical mass estimate for the galaxy. 

Given the \ion{H}{i} masses for the galaxies NGC~59 and ESO~384-016, we are able to derive the \ion{H}{i} gas-to-star mass fractions for these systems (see Table~\ref{ML_ratio1}). The \ion{H}{i} mass of NGC~59 corresponds to $\sim$3\% of its total stellar mass. This galaxy contains the largest amount of \ion{H}{i} gas compared to the other early-type dwarfs in the Scl group \citep{Bou2005,Bea2006}. The relatively large \ion{H}{i} content of NGC~59 together with the ionized gas in the galaxy center \citep{Ski03} suggests that this galaxy cannot be classified as a genuine dS0 galaxy. Instead, this detection supports the claim of \citet{Bou2005} that this dwarf should rather be classified as a mixed-type dS0/Im galaxy as it exhibits the characteristics of both early and late-type galaxies. 

The galaxy ESO~384-016 shows a larger gas fraction to that detected for the star-forming galaxy NGC~59. This correlates with the \ion{H}{i} mass to $B$-band luminosity ratios measured by \citet{Bea2006} where the lower \ion{H}{i} gas fraction was measured for NGC~59. These were found to be $M_{\ion{H}{i}}/L_B = 0.21$ for ESO~384-016 while NGC~59 has $M_{\ion{H}{i}}/L_B = 0.07$. ESO~384-016 represents one of four mixed-type dwarfs in the Cen~A group \citep{Bou2007}. Only two of the mixed-type dwarfs (including ESO~384-016) have been detected in \ion{H}{i}. These galaxies show $M_{\ion{H}{i}}/L_B$ ratios similar to that found for mixed-morphology dwarfs in the Local Group \citep{Ger1999,Bou2006}. In addition, the \ion{H}{i} distribution of ESO~384-016 shows an eastern extension of the gas \citep[see][]{Bea2006}. \citet{Bou2007} suggest that the \ion{H}{i} distribution of this galaxy is a result of mild ram pressure exerted by an intergalactic medium of density $\rho_{\rm IGM}\sim 10^{-3}\rm cm^{-3}$. 


The upper \ion{H}{i} mass limit for NGC~5206 was used to estimate the gas fraction of this dwarf. The stellar mass is the main contributor to the baryonic mass of this galaxy. The \ion{H}{i} gas content of NGC~5206 is found to be less than 0.1\% of the total stellar mass.

\section{Summary and Conclusions}

Deep NIR $J$-, $H$- and $K_s$-band imaging were obtained for a sample of six early-type dwarf galaxies in the LV ($D\la10\rm~Mpc$). The galaxies are detected down to a SB limit of $\mu\simeq24\rm~mag~arcsec^{-2}$ in the $J$- and $H$-bands, and $\mu\simeq23\rm~mag~arcsec^{-2}$ in the $K_s$-band. The low SB galaxies LEDA~166099 and UGCA~200 are detected in the NIR for the first time. A detailed NIR photometric study was conducted for each of the dwarf galaxies to explore the properties and various characteristics of the old stellar component of the galaxy.

The deep NIR observations allow for an accurate measure of the total magnitudes of even the faintest galaxies in our sample. The total magnitudes of the three brightest ($M_B\la-15.5\rm~mag$) galaxies NGC~3115~DW01, NGC~59 and NGC~5206 were compared to those obtained by 2MASS. For these galaxies, we find that 2MASS can underestimate the magnitudes by up to $\la0.5\rm~mag$. The remaining galaxies in our sample were not detected by 2MASS. These findings highlight the selection biases faced when using photometric data from the 2MASS galaxy survey. The structure of the underlying stellar component was determined by fitting an analytical function to the one-dimensional light profile of the galaxy. An exponential law was fit to the stellar disk of those galaxies hosting a nucleus. The effective radius $r_e$ and corresponding SB $\mu_e$ associated with the disk component were derived for these galaxies. The light profiles of the non-nucleated galaxies were best modeled by a S\'ersic law. The six dwarf galaxies are found to have similar NIR structure to early-type dwarf systems in the Virgo Cluster. A first indication of the extension of the relations between the structural parameters ($\mu_e$ and $r_e$) and $H$-band luminosity to low luminosities ($M_H\ga-18\rm~mag$) is given by the faintest galaxies in our sample. This parameter space remains largely unexplored due to the lack of NIR data for low SB dwarf galaxies. 

The $J$-$K_s$ colour was derived for the individual dwarf galaxies to explore its variation with the $K_s$-band luminosity of the galaxy. We found that the six dwarfs exhibit almost constant $J$-$K_s$ which is independent of their luminosity. In addition, the nearly constant $J$-$K_s$ colours with luminosity are observed for early-type dwarf galaxies in the Virgo Cluster. These results suggest that the $J$-$K_s$ colour is not a strong tracer of the galaxy metallicity in early-type dwarf systems. The six dwarfs were also found to have typical $B$-$K_s$ colours to those seen in early-type dwarf galaxies which are independent of environment. The correlation between the $B$- and $H$-band luminosities implies that the dwarf galaxies are not strongly affected by dust attenuation so that similar galaxy morphologies are revealed at optical and NIR wavelengths. Finally, the stellar masses of the six galaxies were determined from the $H$-band observations. The dwarf galaxies are found to have stellar masses in the range of $10^8-10^{10}$\MSUN. For the case of NGC~3115~DW01, the stellar mass was compared to its dynamical mass estimate indicating that this galaxy is DM dominated and a possible candidate for hosting a DM halo. The \ion{H}{i} gas-to-star mass fractions were determined for those galaxies in which neutral hydrogen gas has been detected. The galaxies NGC~59 and ESO~384-016 show \ion{H}{i} gas-to-star mass fractions of $\ga3$\% providing further support that these are mixed-type systems rather than pure dwarf elliptical (dE) or lenticular (dS0) galaxies.

\section*{Acknowledgments}

BdS would like to acknowledge P.~V\"ais\"anen and S.~Barway for their useful comments when writing this paper. Support for the research presented in this paper was provided by the South African National Research Foundation. This publication makes use of data products from the Two Micron All Sky Survey (2MASS), which is a joint project of the University of Massachusetts and the Infrared Processing and Analysis Centre/California Institute of Technology, funded by the National Aeronautics and Space Administration and the National Science Foundation. Extensive use of the Goldmine database was made in this paper.

\label{lastpage}

\end{document}